\documentclass[10pt,journal,compsoc]{IEEEtran}

\ifCLASSOPTIONcompsoc
  \usepackage[nocompress]{cite}
\else
  \usepackage{cite}
\fi
\usepackage[numbers,sort&compress]{natbib} 
\usepackage{url}
\usepackage{cite}
\usepackage{fancyhdr}
\usepackage{float}
\usepackage{graphicx}
\usepackage{epsfig}
\usepackage{bm}
\usepackage{subcaption}
\usepackage{amsmath}
\usepackage{mathrsfs}
\usepackage{array}
\usepackage{booktabs}
\usepackage{amssymb}
\usepackage{multicol}
\usepackage{multirow}
\usepackage{graphicx}
\usepackage{epstopdf}
\usepackage{algorithm}
\usepackage{amsthm}
\usepackage{enumitem}
\usepackage{bbm}
\usepackage{graphicx}
\usepackage{epsfig}
\usepackage{color}
\usepackage{bbding}
\usepackage{pifont}
\usepackage{rotating}
\usepackage{wasysym}
\usepackage{adjustbox}
\usepackage{hyperref}

\ifCLASSINFOpdf
\else
\fi
\begin{document}
%
\title{Responsible Diffusion: A Comprehensive Survey on Safety, Ethics, and Trust in Diffusion Models}
%
%
%
%

\author{Kang~Wei,
        Xin~Yuan,
        Fushuo~Huo,
        Chuan~Ma,
        Long~Yuan,
        Songze~Li,
        Ming~Ding,
        Dacheng~Tao
\IEEEcompsocitemizethanks{\IEEEcompsocthanksitem Kang Wei and Songze Li are with the School of Cyber Science and Engineering, Southeast University, Nanjing, 211189, China (e-mail: kang.wei@seu.edu.cn).
\IEEEcompsocthanksitem Xin Yuan and Ming Ding are with Data61, CSIRO, Sydney, Australia.
\IEEEcompsocthanksitem Chuan Ma is with Zhejiang Laboratory, Hangzhou, China.
\IEEEcompsocthanksitem Long Yuan is with the school of Computer Science and Artificial Intelligence, Wuhan University of Technology, Wuhan, China.
\IEEEcompsocthanksitem Fushuo Huo and Dacheng Tao are with the College of Computing and Data Science, Nanyang Technological University, Singapore.
\IEEEcompsocthanksitem Corresponding authors: Fushuo Huo and Xin Yuan.}
}

%
%

\markboth{Journal of \LaTeX\ Class Files,~Vol.~14, No.~8, August~2015}%
{Shell \MakeLowercase{\textit{et al.}}: Bare Advanced Demo of IEEEtran.cls for IEEE Computer Society Journals}
%



\IEEEtitleabstractindextext{%
\begin{abstract}
Diffusion models (DMs) have been investigated in various domains due to their ability to generate high-quality data, thereby attracting significant attention.
However, similar to traditional deep learning systems, there also exist potential threats to DMs. 
To provide advanced and comprehensive insights into safety, ethics, and trust in DMs, this survey comprehensively elucidates its framework, threats, and countermeasures. 
Each threat and its countermeasures are systematically examined and categorized to facilitate thorough analysis.
Furthermore, we introduce specific examples of how DMs are used, what dangers they might bring, and ways to protect against these dangers.
Finally, we discuss key lessons learned, highlight open challenges related to DM security, and outline prospective research directions in this critical field.
This work aims to accelerate progress not only in the technical capabilities of generative artificial intelligence but also in the maturity and wisdom of its application.
\end{abstract}

\begin{IEEEkeywords}
Diffusion models, text-to-image, privacy, robustness, safety, fairness, copyright and
truthfulness.
\end{IEEEkeywords}}

\maketitle

\IEEEdisplaynontitleabstractindextext

%
\IEEEpeerreviewmaketitle

\ifCLASSOPTIONcompsoc
\IEEEraisesectionheading{\section{Introduction}\label{sec:introduction}}
\else
\section{Introduction}
\label{sec:introduction}
\fi

%
%
%
%
\IEEEPARstart{R}{ecently}, generative artificial intelligence (GAI) has attracted significant attention due to its ability to generate high-quality data \cite{Chen2024Opportunities}.
Different from discrimination tasks, GAI focuses on generating new samples from the learned distribution, thereby demonstrating remarkable creativity. As one of the famous generative models, diffusion models (DMs) are widely used in the image and video generation domains \cite{Croitoru2023Diffusion}.
In general, DMs consist of two main processes, i.e., forward and reverse processes. 
The forward (diffusion) process gradually adds noise to the original data to diffuse the data distribution into the standard Gaussian distribution. 
The reverse (generative) process employs a deep neural network, which is often a UNet \cite{Christopher2023A}, to reverse the diffusion, reconstructing the data from the Gaussian noise.
With its impressive potential, DMs have been investigated in various domains, including computer vision \cite{Croitoru2023Diffusion}, natural language processing (NLP) \cite{Chen2023TextDiffuser}, audio processing \cite{Lemercier2024Diffusion}, 3D generation~\cite{shi2024mvdream}, bioinformatics \cite{Gao2024Diffusion}, and time series tasks \cite{Marcel2023Predict}.

DMs have shown great potential in various fields, they not only face traditional privacy and security threats similar to traditional deep learning (DL) systems~\cite{Ma2023Trusted}, but also include specific challenges~\cite{Zhou2022Adversarial}.
For instance, attackers may leverage the memorization capabilities of DMs to infer data privacy.
These attacks are usually related to privacy leakage or intellectual property theft, in which adversaries aim to extract sensitive information or training data from a model or its outputs.
From the aspect of robustness, there exist several threats, such as adversarial attacks.
Adversarial attacks are also recognized as harmful to DL systems, in which an attacker intentionally perturbs input data to deceive a model into making incorrect predictions or classifications. 
These attacks exploit vulnerabilities in DL models by finding small, often imperceptible changes to inputs that lead to significant errors in the model's output.
In addition to privacy and robustness issues, DMs are also vulnerable to safety risks, e.g., backdoor attacks~\cite{Chou2023How,Chen2023TrojDiff,Zhai2023Text,Chou2023VillanDiffusion}. 
Such threats arise when hidden backdoors are embedded into DM systems, often due to insufficient control over the training process, such as training on poisoned datasets or using backdoored third-party models~\cite{Li2024Backdoor}.
Fairness remains a classic research topic in data-driven ML algorithms, 
as models may learn and perpetuate biases present in the training data, which can be reflected in their predictions/generation.

Unlike traditional discriminative models that produce classified outputs (e.g., binary decisions), the generated results from GAI could exacerbate specific challenges like: 1) jailbreaks, in which the attacker may perturb the prompt or image input to evade the safety checker and generate unsafe content; 2) data misuse, potentially leading to copyright disputes; 3) hallucination, where generated content may not align with reality. 
Concretely, jailbreak is usually achieved by adversarial inputs, which focuses on generating unsafe contents instead of incorrect results (as done by adversarial attacks).
Copyright aims to protect original human-authored works, which has drawn a lot of attention in DMs \cite{zhang2024on}. 
We can note that there exist various copyright risks in DMs, i.e., model extraction, prompt stealing, and data misuse, which are necessary to be explored. 
Hallucination in DMs is usually defined as truthfulness, which measures whether a generated output accurately corresponds to a set of established facts, specific instructions, and verifiable reality \cite{Xu2023ImageReward}.
\begin{figure*}[tbhp]
\centering
\includegraphics[width=0.78\linewidth]{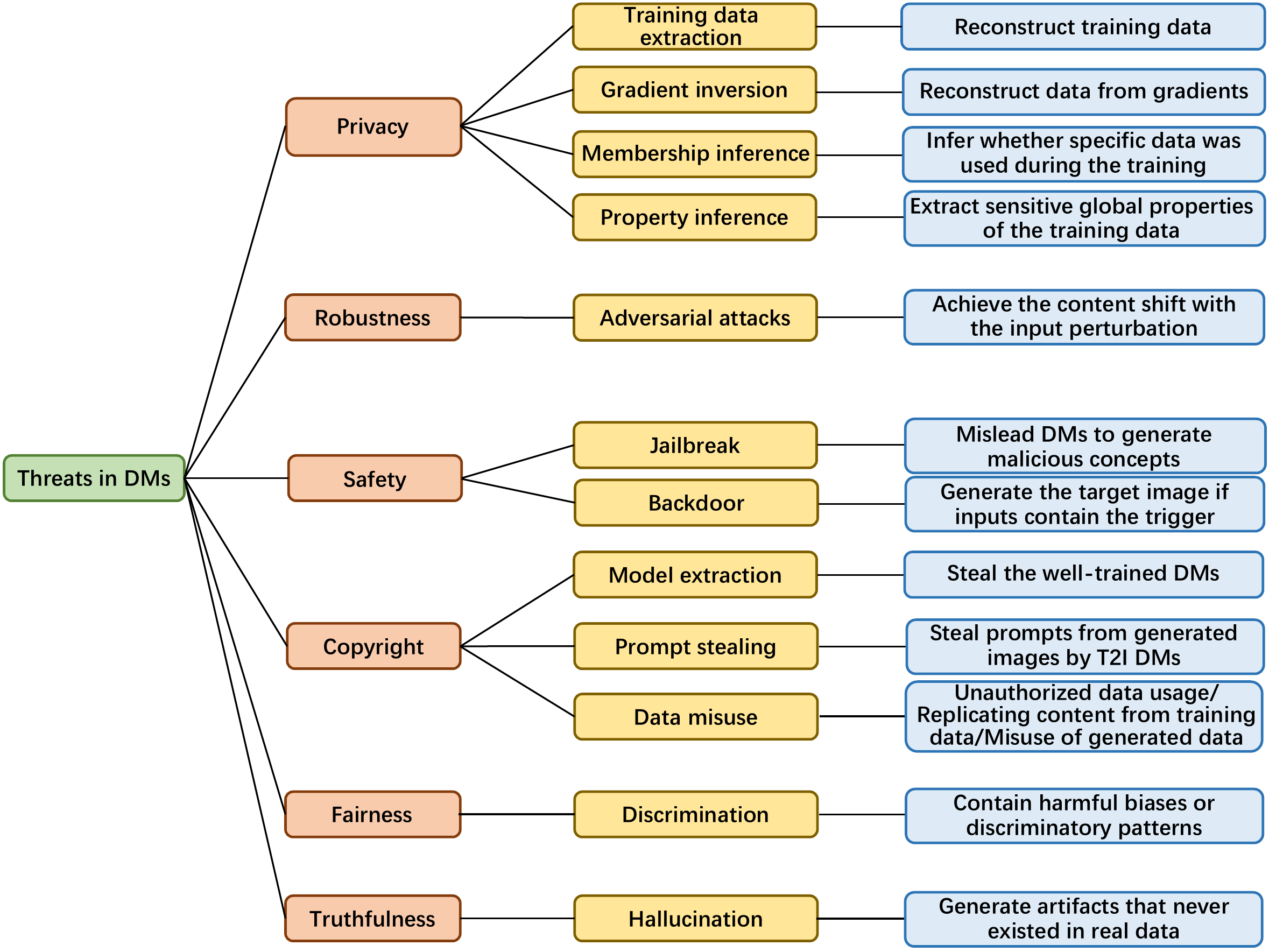}
\caption{Categories of threats in DMs.}
\label{fig:categories of threats}
\end{figure*}

\vspace{-4mm}
Naturally, countermeasures against these threats in DMs are still in early development, focusing on content control, threat removal, and privacy protection.
For content control, 
prompt filter and safety checker are two widely used methods without fine-tuning the model parameters.
Another method is called concept negation, aiming to enable DMs to govern the output space based on concepts.
The design of threat removal is typically tailored to the specific attack.
Taking the backdoor attack as an example, backdoor removal methods can usually be divided into two categories. 
The first category mitigates backdoor attacks with some general methods, such as model fine-tuning and noise perturbation \cite{chew2024defending}.
The second will detect and recover the triggers, and then remove the backdoor with these recovered triggers \cite{an2024how, Wang2025T2IShield,hao2024diffcleanse,mo2024terd}.
For the privacy protection in DMs, the main challenge in classic techniques, e.g., homomorphic encryption (HE), differential privacy (DP)~\cite{9069945, 10073536}, secure multi-party computation (MPC), is to balance efficiency and performance.
In this context, well-designed techniques, e.g., mitigating DMs' memorization, demonstrate tremendous potential.
\begin{table*}
\centering
\caption{Existing surveys on threats and defenses in DMs. The symbols \CIRCLE, \LEFTcircle~and \Circle~indicate full coverage, partial coverage, and no coverage of the corresponding content, respectively.}\label{tab:survey}
\resizebox{\linewidth}{!}{
\begin{tabular}{|m{0.5cm}<{\centering}|m{1.4cm}<{\centering}|m{1.0cm}<{\centering}|m{1.8cm}<{\centering}|m{1.8cm}<{\centering}|m{1.6cm}<{\centering}|m{1.4cm}<{\centering}|m{1.1cm}
<{\centering}|m{1.2cm}<{\centering}|m{1.0cm}<{\centering}|m{1.4cm}<{\centering}|m{1.2cm}<{\centering}|m{1.5cm}<{\centering}|}
\hline
\multirow{3}*{\shortstack{Ref.}}&\multicolumn{2}{c|}{Application Category}&\multicolumn{3}{c|}{Privacy}& Robustness & \multicolumn{2}{c|}{Safety}& Fairness&\multicolumn{2}{c|}{Copyright}& Truthfulness\\
\cline{2-13}
&Image Generation& Text-to-Image&Training Data Extraction& Data Reconstruction& Membership Inference&Adversarial Attack & Backdoor&Jailbreak&Bias&Model Extraction& Prompt Stealing&Hallucina- tion\\
\hline\hline
\cite{zhang2024trustworthy}& \Circle& \CIRCLE& \CIRCLE&\Circle &\CIRCLE&\CIRCLE & \CIRCLE& \CIRCLE &\CIRCLE&\Circle&\Circle&\Circle\\
\hline
\cite{truong2024attacks} &\CIRCLE&\LEFTcircle&\Circle&\Circle&\CIRCLE&\CIRCLE&\CIRCLE&\Circle&\Circle&\Circle&\Circle&\Circle\\
\hline
\cite{ZHANG2025A} &\Circle&\CIRCLE&\Circle&\Circle&\Circle&\CIRCLE&\Circle&\Circle&\Circle&\Circle&\Circle&\Circle\\
\hline\hline
Ours &\CIRCLE&\CIRCLE&\CIRCLE&\CIRCLE&\CIRCLE&\CIRCLE&\CIRCLE&\CIRCLE&\CIRCLE&\CIRCLE&\CIRCLE&\CIRCLE\\
\hline
\end{tabular}
}
\end{table*}

Although there have already existed survey papers on threats and defenses of DMs, they mainly focus on classic attacks, such as membership inference and backdoor, failing to review attack and defense methods comprehensively.
Specifically, Zhou \textit{et al.} \cite{zhang2024trustworthy} first surveyed security risks in DMs, but focused only on text-to-image (T2I) DMs.
Truong \textit{et al.} \cite{truong2024attacks} presented a comprehensive analysis of the security and privacy landscape of DMs, focusing on three specific attack and defense methods for DMs, i.e., backdoor, inference, and adversarial attacks, but fairness, copyright, and truthfulness are unexplored.
The work in \cite{ZHANG2025A} only reviewed adversarial attacks and defenses on T2I DMs, which have inherent limitations.
In order to help researchers better understand this topic and pursue future research work, a comprehensive study is necessary. 
\begin{table}
\centering
\scriptsize
\caption{Summary of important acronyms}
\begin{tabular}{|c|c|}
\hline
\textbf{Acronym}&\textbf{Definition}\\
\hline
\hline
DM/DMs&Diffusion Model/Diffusion Models\\
\hline
DDPM & Denoising Diffusion Probabilistic Model\\
\hline
DDIM & Denoising Diffusion Implicit Model\\
\hline
DP& Differential Privacy\\
\hline
HE & Homomorphic Encryption\\
\hline
MPC & Multi-party Computation\\
\hline
SGMs&Core-based Generative Models\\
\hline
SDEs & Stochastic Differential Equations\\
\hline
SGD & Stochastic Gradient Descent\\
\hline
KL&Kullback-Leibler\\
\hline
LLM & Large Language Model\\
\hline
MIA & Membership Inference Attack\\
\hline
GANs & Generative Adversarial Networks\\
\hline
LDM & Latent Diffusion Model\\
\hline
SLD & Safe Latent Diffusion\\
\hline
T2I & Text-to-Image \\
\hline
NSFW  & Not Safe For Work \\
\hline
IBA & Invisible Backdoor Attack\\
\hline
OOD & Out-of-Distribution\\
\hline
\end{tabular}
\label{tab:Acronyms}
\end{table}
To fill this gap, we provide a systematic and comprehensive overview of state-of-the-art (SOTA) research studies on attacks and countermeasures in DMs across six primary aspects, i.e., privacy, robustness, safety, fairness, copyright, and truthfulness, as shown in Fig. \ref{fig:categories of threats}.
The main contributions are summarized as follows:
\begin{itemize}[leftmargin=1.2em]
    \item We provide readers with the necessary background knowledge of different types of DMs. We compare these DMs from the aspects of the forward process, the backward process, and the objective. 
    \item We investigate security and privacy issues on DMs across six main groups, including privacy, robustness, safety, fairness, copyright, and truthfulness. Each attack is categorized further into sub-groups based on the corresponding purposes.
    \item We survey various countermeasures for DM-targeted attacks. These countermeasures can be divided into several categories: content control, threat removal, and privacy protection.
    Each countermeasure can be utilized to defend against one or more security and privacy issues on DMs.
    \item We discuss open challenges in this topic and propose promising research directions to improve the security aspect of DMs and their applications.
\end{itemize}

To further emphasize our contributions of this paper, we present a comparison between our work and existing DM-related surveys in Tab. \ref{tab:survey}.
In addition, the summary of important acronyms is shown in Tab. \ref{tab:Acronyms}.
\section{Preliminaries}\label{sec:DMs}
\begin{figure}[t]
\centering
\includegraphics[width=1.0\linewidth]{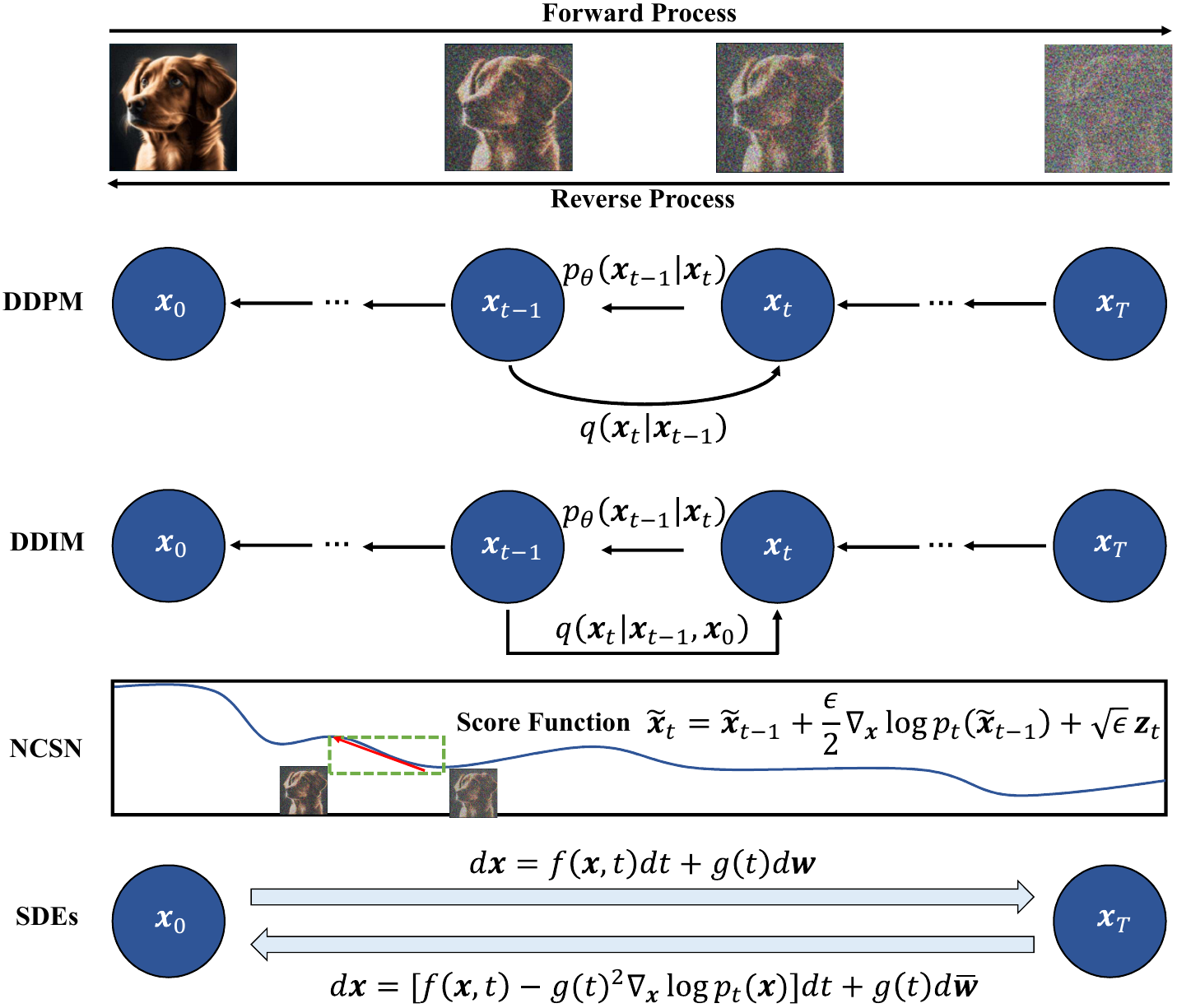}
\caption{Overall framework of four
types of DMs with different forward/backward processes or objective functions.}
\label{fig:DM_framework}
\end{figure}

DMs are a powerful class of probabilistic generative models that learn to produce high-quality samples by methodically reversing a fixed process where data is gradually degraded into pure noise.
In this section, we will provide a brief introduction to the four classic DMs, i.e., denoising diffusion probabilistic model (DDPM), denoising diffusion implicit model (DDIM), score-based generative model (SGM) and stochastic differential equations (SDEs), as shown in Fig. \ref{fig:DM_framework}.
\subsection{Unconditional DMs}
\subsubsection{DDPM and DDIM} 
\noindent\textbf{DDPM:} The DDPM framework is defined by two Markov chains: a fixed forward process that incrementally adds noise to data, and a learned reverse process that generates data by reversing the diffusion.
Given a data distribution $\boldsymbol{x}_{0}\sim q(\boldsymbol{x}_{0})$, the forward process generates a sequence of latent variables, typically denoted as $\boldsymbol{x}_{1}, \boldsymbol{x}_{2}, \ldots \boldsymbol{x}_{T}$ with transition kernel $q(\boldsymbol{x}_{t}|\boldsymbol{x}_{t-1})$.
By applying the chain rule of probability and the Markov property, we factorize the joint distribution of $\boldsymbol{x}_{1}, \boldsymbol{x}_{2}, \ldots \boldsymbol{x}_{T}$ conditioned on $\boldsymbol{x}_{0}$, denoted as $q(\boldsymbol{x}_{1},\ldots,\boldsymbol{x}_{T}|\boldsymbol{x}_{0})$.
The forward process, a.k.a., the diffusion process, is a Markov chain that systematically incorporates Gaussian noise into the data based on a fixed variance schedule $\{\beta_1, \ldots, \beta_T\}$, obtained as
\begin{equation}
q(\boldsymbol{x}_{1},\ldots,\boldsymbol{x}_{T}|\boldsymbol{x}_{0}) = \prod_{t=1}^{T}q(\boldsymbol{x}_{t}|\boldsymbol{x}_{t-1}),
\end{equation}
where
$q(\boldsymbol{x}_{t}|\boldsymbol{x}_{t-1}) = \mathcal{N}\left(\boldsymbol{x}_{t}; \sqrt{1-\beta_{t}}\boldsymbol{x}_{t-1}, \beta_{t}\boldsymbol{I}\right)$.

To produce new samples, DDPM starts by drawing noise from a simple prior distribution (typically Gaussian) and then iteratively denoise it by running a learned reverse Markov chain, i.e.,
\begin{equation}
p_{\boldsymbol{\theta}}(\boldsymbol{x}_{0:T}) := p(\boldsymbol{x}_{T})\prod_{t=1}^{T}p_{\boldsymbol{\theta}}(\boldsymbol{x}_{t-1}|\boldsymbol{x}_{t}),
\end{equation}
where 
$p_{\boldsymbol{\theta}}(\boldsymbol{x}_{t-1}|\boldsymbol{x}_{t}):=\mathcal{N}\left(\boldsymbol{x}_{t-1}; \boldsymbol{\mu}_{\boldsymbol{\theta}}(\boldsymbol{x}_{t}, t), \boldsymbol{\Sigma}_{\boldsymbol{\theta}}(\boldsymbol{x}_{t}, t)\right)$. Here,
$\boldsymbol{\theta}$ is the learnable parameter, $\boldsymbol{\mu}_{\boldsymbol{\theta}}(\boldsymbol{x}_{t}, t)$ and $\boldsymbol{\Sigma}_{\boldsymbol{\theta}}(\boldsymbol{x}_{t}, t)$ are learnable mean and variance of the reverse Gaussian kernels, respectively.
The training process optimizes the variational lower bound of the negative log-likelihood as
\begin{equation}
\begin{aligned}
L &\triangleq\mathbb{E}[-\log p_{\boldsymbol{\theta}}(\boldsymbol{x}_{0})] \leq \mathbb{E}_{q}\left[-\log\frac{p_{\boldsymbol{\theta}}(\boldsymbol{x}_{0:T})}{q(\boldsymbol{x}_{1:T}|\boldsymbol{x}_{0})}\right]\\
&=\mathbb{E}_{q}\left[-\log p(\boldsymbol{x}_{T})-\sum_{t\geq 1}\log \frac{p_{\boldsymbol{\theta}}(\boldsymbol{x}_{t-1}|\boldsymbol{x}_{t})}{q(\boldsymbol{x}_{t}|\boldsymbol{x}_{t-1})}\right].
\end{aligned}
\end{equation}

Efficient training can be attained by optimizing random terms of $L$ via stochastic gradient descent (SGD).
Variance reduction yields better performance by rewriting $L$ as:
\begin{equation}
\begin{aligned}
&\mathbb{E}_{q}\Bigg{[}\underbrace{D_{KL}(q(\boldsymbol{x}_{T}|\boldsymbol{x}_{0})\Vert p(\boldsymbol{x}_{T}))}_{L_T}\underbrace{-\log p_{\boldsymbol{\theta}}(\boldsymbol{x}_{0}|\boldsymbol{x}_{1})}_{L_0}\\
&\qquad \quad +\sum_{t\geq1} \underbrace{D_{KL}(q(\boldsymbol{x}_{t-1}|\boldsymbol{x}_{t},\boldsymbol{x}_{0})\Vert p_{\boldsymbol{\theta}}(\boldsymbol{x}_{t-1}|\boldsymbol{x}_{t}))}_{L_{t-1}}\Bigg{]}.
\end{aligned}
\end{equation}

\noindent\textbf{DDIM:}
As an efficient class of iterative implicit probabilistic models, DDIM was proposed in~\cite{song2021denoising}.
The key idea of DDIM is to develop non-Markovian inference processes, leading to the same surrogate objective function as DDPM.
The generative process can be expressed as:
\begin{equation}
\begin{aligned}
\boldsymbol{x}_{t-1} &= \sqrt{\alpha_{t-1}}\left(\frac{\boldsymbol{x}_{t}-\sqrt{1-\alpha_{t}}\epsilon_{\boldsymbol{\theta}}^{(t)}(\boldsymbol{x}_{t})}{\sqrt{\alpha_{t}}}\right) \\ & \qquad\qquad + \sqrt{1-\alpha_{t-1}-\sigma_{t}^{2}}\epsilon_{\boldsymbol{\theta}}^{(t)}(\boldsymbol{x}_{t})+\sigma_{t} \epsilon_{t},
\end{aligned}
\end{equation}
where $\epsilon_{t} \sim \mathcal{N}(0, \boldsymbol{I})$ denotes standard Gaussian noise independent of $\boldsymbol{x}_{t}$, and $\alpha_{0} =1 $. 
Adjusting $\sigma_{t}$ yields different generative processes using the same $\epsilon_{\boldsymbol{\theta}}$, eliminating the need for retraining. In particular, when $\sigma_{t} = \sqrt{(1-\alpha_{t-1})/(1-\alpha_{t})} \sqrt{1-\alpha_{t}/\alpha_{t-1}}$, $\forall t$, the forward process is Markovian, and the generative process is a DDPM.
\subsubsection{SGM}
The key idea of SGM is to perturb data using a sequence of increasingly strong Gaussian noise levels, while simultaneously learning the score functions of all noisy data distributions by training a deep neural network model conditioned on noise levels \cite{Yang2019Generative}.
Langevin dynamics generate samples from a target distribution $p(\boldsymbol{x})$ using sorely its score function $\nabla_{\boldsymbol{x}} \log p(\boldsymbol{x})$. 
Starting from an initial point $\boldsymbol{\tilde{x}} \sim \pi (\boldsymbol{x})$, where $\pi$ is a prior distribution, and given a fixed step size $\epsilon > 0$, the method updates recursively according to the following
\begin{equation}
\begin{aligned}
\boldsymbol{\tilde{x}}_{t} = \boldsymbol{\tilde{x}}_{t-1} + \frac{\epsilon}{2} \nabla_{\boldsymbol{x}} \log p(\boldsymbol{\tilde{x}}_{t-1})+\sqrt{\epsilon} \boldsymbol{z}_{t},
\end{aligned}
\end{equation}
where $\boldsymbol{z}_{t} \sim \mathcal{N}(0, \boldsymbol{I})$.
As $\epsilon \rightarrow 0 $ and $T \rightarrow \infty$, the distribution of $\boldsymbol{\tilde{x}}_{T}$ converges to $p(\boldsymbol{x})$, indicating that under certain regularity assumptions, $p(\boldsymbol{x})$ is an exact sample from $p(\boldsymbol{x})$.


It can be noted that $s_{\boldsymbol{\theta}}(\boldsymbol{x}, \sigma)$ is called a Noise Conditional Score Network (NCSN).
Empirically, both sliced score matching and denoising score matching can be used to train NCSN.
For a given $\sigma$, 
combining this equation for all $\sigma \in \{ \sigma_{i}\}_{i=1}^{L}$ to get one unified objective as
\begin{equation}
\begin{aligned}
&\ell (\boldsymbol{\theta}; \{ \sigma_{i}\}_{i=1}^{L}) \triangleq \frac{1}{L} \sum_{i=1}^{L} \lambda(\sigma_{i})\ell(\boldsymbol{\theta};\sigma_{i}),
\end{aligned}
\end{equation}
where $\lambda(\sigma_{i}) > 0$ is a coefficient function and
\begin{equation}
\begin{aligned}
&\ell (\boldsymbol{\theta}; \sigma) \triangleq \\
&\quad \frac{1}{2} \mathbb{E}_{p_{\text{data}}(\boldsymbol{x})}\mathbb{E}_{\boldsymbol{\tilde{x}}\sim \mathcal{N}(\boldsymbol{x},\sigma^{2}\boldsymbol{I})}\left[\left\Vert s_{\boldsymbol{\theta}}(\boldsymbol{\tilde{x}}, \sigma) + (\boldsymbol{\tilde{x}}-\boldsymbol{x})/\sigma^2\right\Vert^2_{2}\right].
\end{aligned}
\end{equation}
\subsubsection{SDEs}
Score SDEs gradually perturb data into noise via a diffusion process~\cite{song2021scorebased}:
${\rm{d}} \boldsymbol{x} = {\rm{f}}(\boldsymbol{x}, t) {\rm{d}}t+g(t){\rm{d}}\boldsymbol{w}$,
where $\boldsymbol{w}$ denotes the standard Wiener process (a.k.a., Brownian motion), ${\rm{f}}(\cdot, t)$ is the drift term of $\boldsymbol{x}(t)$, and $g(\cdot)$ is the diffusion coefficient. 
The reverse of this process is also a diffusion process, evolving backwards in time according to:
\begin{equation}
\begin{aligned}
{\rm{d}} \boldsymbol{x} = [{\rm{f}}(\boldsymbol{x}, t)-g(t)^{2}\nabla_{\boldsymbol{x}}\log p_{t}(\boldsymbol{x})] {\rm{d}}t+g(t){\rm{d}}\overline{\boldsymbol{w}},
\end{aligned}
\end{equation}
where $\overline{\boldsymbol{w}}$ is a standard Wiener process in reverse time, and ${\rm{d}}t$ indicates a backward time step.
If the score function $\nabla_{\boldsymbol{x}}\log p_{t}(\boldsymbol{x})$ is available for all $t$, we can derive and simulate the above reverse SDE to generate samples from $p_{0}$.
This score function can be approximated by training a time-dependent score-based model $s_{\boldsymbol{\theta}}(\boldsymbol{x},t)$ on noisy samples using score matching, as given by
\begin{equation}
\begin{aligned}
\boldsymbol{\theta}^{\star} & = \arg\min_{\boldsymbol{\theta}} \mathbb{E}_{t} \big{\{} \lambda(t) \mathbb{E}_{\boldsymbol{x}(0)}\mathbb{E}_{\boldsymbol{x}(t)|\boldsymbol{x}(0)}\\
&\qquad[\Vert s_{\boldsymbol{\theta}}(\boldsymbol{x}(t),t)-\nabla_{\boldsymbol{x}}\log p_{0t}(\boldsymbol{x}(t)|\boldsymbol{x}(0))\Vert^{2}_{2}] \big{\}}.
\end{aligned}
\end{equation}
Here $\lambda(t) > 0$ is a positive weighting function, $t$ is uniformly sampled over $[0, T]$,
$\boldsymbol{x}(0)\sim p_{0}(\boldsymbol{x}(t))$ and $\boldsymbol{x}(t) \sim p_{0t}(\boldsymbol{x}(t)|\boldsymbol{x}(0))$.
\subsection{Conditional DMs}
Conditional DMs are a class of generative models that extend the capabilities of DMs by allowing them to generate data conditioned on additional information~\cite{ko2024stochastic}, such as generating images based on textual descriptions or other attributes. 
Specifically, conditional DMs are prevalently leveraged for conditional generations by modifying the U-Net architecture to incorporate the conditions into the network~\cite{Christopher2023A}.
As one of its most widely used implementations, latent diffusion model (LDM) is a typical text-to-image (T2I) DM \cite{Rombach2022High}.
LDM introduced to project text prompt $y$ with a domain specific encoder $\tau_{\boldsymbol{\theta}}$ and a cross-attention layer, which is expressed as $\tau_{\boldsymbol{\theta}}(y) \in \mathbb{R}^{M\times d_{\tau}}$ and
\begin{equation}
\text{Attention}(Q, K, V) = \text{softmax}\left(\frac{QK^{\top}}{\sqrt{d}}\right) \cdot V,
\end{equation}
respectively, where $Q=W_{Q}^{(i)}\cdot\rho_{i}(\boldsymbol{z}_{t})$, $K=W_{K}^{(i)}\cdot \tau_{\boldsymbol{\theta}}(y)$, $V=W_{V}^{(i)}\cdot \tau_{\boldsymbol{\theta}}(y)$, $\rho_{i}(\boldsymbol{z}_{t})$ is a (flattened) intermediate representation of the U-Net, $W_{Q}^{(i)}$, $W_{K}^{(i)}$ and $W_{V}^{(i)}\cdot \tau_{\boldsymbol{\theta}}(y)$ are learnable parameters, $\boldsymbol{z}_{t}$ is a latent representation encoded from $\boldsymbol{x}$.
Based on image-conditioning pairs, LDM learns the conditional generative model via
\begin{equation}
L_{\text{LDM}} := \mathbb{E}_{\boldsymbol{x},y,\epsilon\sim \mathcal{N}(0,1),t}\left[\left\Vert \epsilon-\epsilon_{\boldsymbol{\theta}}(\boldsymbol{z}_{t},t,\tau_{\boldsymbol{\theta}}(y))\right\Vert_{2}^{2}\right],
\end{equation}
where $\epsilon_{\boldsymbol{\theta}}(z_{t},t,\tau_{\boldsymbol{\theta}}(y))$ is the denoising autoencoder.

Subsequently, advances in T2I DMs, such as Uni-ControlNet~\cite{Zhao2023UniControlNet}, GLIGEN \cite{Li2023GLIGEN} and DesignDiffusion \cite{Wang2025DesignDiffusion}, have been proposed, paving the way for adding spatial controls to large pretrained DMs.
Uni-ControlNet is a unified framework that enables flexible, simultaneous integration of both local controls (e.g., edge maps, depth maps, segmentation masks) and global controls (e.g., CLIP image embeddings) within a single model. 
GLIGEN enhances pre-trained T2I DMs by freezing all of its weights and introduces new trainable layers with a gated mechanism to incorporate grounding information.
DesignDiffusion incorporates a unique character embedding extracted from visual text to augment input prompts, combined with a character localization loss for stronger generation supervision, as well as a self-play direct preference optimization (DPO) fine-tuning strategy to enhance visual text synthesis performance.
OpenAI also developed its own T2I DM, i.e., DALL·E, which is integrated with ChatGPT for richer prompt understanding, improved detail, and reduced artifacts. 
Google also introduced Imagen that emphasizes textual fidelity and photorealistic quality through its unique architecture.

As conditional DMs are widely used in commercial applications and have attracted significant community interest due to their powerful guided generation capabilities, it is important to study the additional risks, such as adversarial input \cite{Yu2024Step} and data poisoning \cite{Zhai2023Text}.
\section{Attacker's Capabilities}
\noindent\textbf{Black-box attack:} 
Under the black-box setting, the most difficult and realistic scenario, attackers can observe the final output of the target DMs, i.e., the generated images, and have no access to the model parameters, hyperparameters, architecture, or other internal details. 

\noindent\textbf{Gray-box attack:} 
During the sampling process, DMs iteratively denoise an image over multiple steps, producing numerous intermediate outputs. 
In a gray-box setting, attackers exploit this by manipulating the target model's generation process to extract the intermediate output of the U-Net
model at each timestep, while still lacking access to parameters or full internal details.

\noindent\textbf{White-box attack:} 
In this setting, attackers have unrestricted access to the target model's parameters, architecture, source codes, and other internal details. Such conditions are common in the open-source community, where source codes, models’ information, and pretrained model checkpoints are publicly available and easily accessible.
\section{Threats}
In this section, we review existing works on threats to DMs across six aspects, i.e., privacy, robustness, safety, copyright, fairness, and truthfulness, each of which is further divided into classes based on the attacker’s objectives.
\begin{table*}[h]
\caption{Taxonomy of privacy risks in DMs.}
\centering
\scriptsize
\begin{tabular}{|m{1.3cm}<{\centering}|m{0.3cm}<{\centering}|m{1.3cm}<{\centering}|m{2.3cm}<{\centering}|m{5.4cm}<{\centering}|m{5.0cm}<{\centering}|}
\hline
\textbf{Category}& \textbf{Ref.}& \textbf{Attacker's knowledge}& \textbf{Target Models}& \textbf{Effectiveness}&\textbf{Limitations}\\
\hline\hline
\multirow{5}*{\shortstack{Training \\ data \\  extraction}} &\cite{Nicolas2023Extracting}&  Black-box, White-box & OpenAI-DDPM, DDPM & Extract training samples from DMs&Explainability of regenerating parts of their training datasets\\
\cline{2-6}
&\cite{wen2024detecting}&White-box&DDIM&Able to achieve an AUC of 0.999 and TPR@1\%FPR of 0.988& A tunable threshold is required to be determined \\
\cline{2-6}
&\cite{chen2024extracting}&White-box&DDIM&Highly effective in extracting memorized data&Mainly uses human-labeled data, which is scarce and limits its practical use\\
\cline{2-6}
\cline{1-6}
Gradient inversion&\cite{huang2024gradient} &Grey-box, Black-box&DDPM& Reconstruct high-resolution images from the gradients& Single batch size\\
\cline{1-6}
\multirow{22}*{\shortstack{Membership\\ inference}}
& \cite{wu2022membership} &Black-box&LDM, DALL-E mini &Achieve remarkable attack performance&Requires auxiliary datasets\\
\cline{2-6}
&\cite{Duan2023Are}& Gray-box & SD, LDM, DDPM &Infer the memberships of training samples& Access to the step-wise query results of DMs\\
\cline{2-6}
& \cite{Matsumoto2023Membership} &White-box, Black-box&DDIM&The resilience of DMs to MIA is comparable to that of GANs& When training samples are small\\ 
\cline{2-6}
& \cite{Tang2024Membership} &White-box, Black-box&DDPM&Quantile regression models are trained to predict reconstruction loss distribution for unseen data
&Direct access to the trained model\\
\cline{2-6}
& \cite{kong2024an} &Gray-box&DDPM, SD&Utilize groundtruth trajectory and predicted point to infer memberships&Access
intermediate outputs of DMs\\
\cline{2-6}
& \cite{zhai2024membership} &Gray-box&SD v1-4&Estimate the gap between conditional image-text likelihood and images-only likelihood & Evaluations under the pretraining setting are insufficient\\
\cline{2-6}
& \cite{fu2024model} &Gray-box&LDM, DDPM&Leverage the intrinsic generative priors within the DMs& Lack sufficient theoretical
evidence\\
\cline{2-6}
& \cite{Li2024Unveiling} &Gray-box&DDIM&First investigate the structural changes during the diffusion process& Limited to structure-based MIA\\
\cline{2-6}
& \cite{Zhang2024Generated} &Black-box&DDPM, DDIM, FastDPM, VQGAN, LDM, CC-FPSE&Generalize MIA against various generative models including GANs, VAEs, IFs, and DDPMs& Requires auxiliary datasets\\
\cline{2-6}
& \cite{li2024black} &Black-box&DDIM, Diffusion Transformer, SD, DALL-E 2&Only require access to the variation API of the model&Requirement for a moderate diffusion
step $t$ in the variation API\\
\cline{2-6}
& \cite{Pang2025Black} &Black-box& SD, DDIM&A score-based MIA tailored for modern DMs, which operates in a black-box setting& Rely on a captioning model previously fine-tuned with an auxiliary dataset\\
\hline
Property inference &\cite{hu2024prisampler}&Black-box&TabDDPM, DDPM, SMLD, VPSDE, VESDE& DMs and their sampling methods are susceptible property inference attacks& Requires one part
of the training datasets as auxiliary datasets\\
\hline
\end{tabular}
\label{tab:privacy_issues}
\end{table*}
\begin{table*}[h]
\caption{Taxonomy of defense methods against privacy issues in DMs.}
\centering
\scriptsize
\begin{tabular}{|m{1.2cm}<{\centering}|m{0.3cm}<{\centering}|m{3.7cm}<{\centering}|m{2.5cm}<{\centering}|m{8.3cm}<{\centering}|}
\hline
\textbf{Threats}& \textbf{Ref.}&\textbf{Key Methods}& \textbf{Target Models}& \textbf{Effectiveness}\\
\hline\hline
\multirow{10}*{\shortstack{Training \\ data \\  extraction}}&\multirow{5}*{\shortstack{\cite{Nicolas2023Extracting}}}&DP-SGD&OpenAI-DDPM, DDPM&Cause the training to fail\\
\cline{3-5}
&&Deduplicating&OpenAI-DDPM, DDPM&Reducing memorization revealed a stronger correlation between training data extraction and duplication rates in models trained on larger-scale datasets\\
\cline{3-5}
&&Auditing with Canaries&OpenAI-DDPM, DDPM&When auditing less leaky models, however, canary exposures computed from a single training might underestimate the true data leakage\\
\cline{2-5}
&\multirow{5}*{\shortstack{\cite{wen2024detecting}}}&Straightforward method to detect trigger tokens&SD & Ensure a more consistent alignment between prompts and generations\\
\cline{3-5}
&&Inference-time mitigation method&SD&\multirow{2}*{\shortstack{Successfully mitigate the memorization effect and offer a more favorable\\ CLIP score trade-off compared to RTA}}\\
\cline{3-4}
&&Training-time mitigation method&SD & \\
\cline{2-5}
&\cite{Ren2024Unveiling} &Detection and mitigation&SD v1.4, SD v2.0& Adding almost no computational cost, maintaining fast training and inference, significantly reducing Similarity Scores from 0.7 to 0.25-0.3\\
\cline{1-5}
\multirow{11}*{\shortstack{Membership\\ inference}} &\multirow{5}*{\shortstack{\cite{Duan2023Are}}}
&Cutout&\multirow{5}*{\shortstack{DDPM}}&\multirow{2}*{\shortstack{ASR and AUC drop to some extent}}\\
\cline{3-3}
&&RandomHorizontalFlip&&\\
\cline{3-3}
\cline{5-5}
&&RandAugment&&\multirow{3}*{\shortstack{Fail to converge}}\\
\cline{3-3}
&&DP-SGD&&\\
\cline{3-3}
&&$\ell_{2}$-regularization&&\\
\cline{2-5}
&\multirow{2}*{\shortstack{\cite{tran2024dual}}}&DualMD&SD v1.5, DDPM&More effective for T2I DMs\\
\cline{3-5}
&&DistillMD& SD v1.5, DDPM &Better suited for unconditional DMs\\
\cline{2-5}
&\multirow{3}*{\shortstack{\cite{Luo2025Privacy}}}&MP-LoRA&SD v1.5&Reduces the ASR to near-random performance while impairing LoRA’s image generation capability\\
\cline{3-5}
&&SMP-LoRA&SD v1.5&Effectively protects membership privacy while maintaining strong image generation quality\\
\cline{1-5}
Property inference &\cite{hu2024prisampler}&PriSampler&TabDDPM &Lead adversaries to infer property proportions that approximate the predefined values specified by model owners \\
\hline
\end{tabular}
\label{tab:privacy_defense}
\end{table*}
\subsection{Privacy}\label{subsec:privacy}


When DMs are trained on sensitive or private data, they pose significant privacy risks due to their potential to inadvertently expose confidential information from the training data. The core vulnerability lies in the model's tendency to memorize specific details from training samples and subsequently reproduce these details in generated outputs, creating pathways for privacy leakage \cite{gu2023memorization,somepalli2023understanding}.
Privacy issues in DMs can be mainly categorized into four types: 1) training data extraction, 2) gradient inversion, 3) membership inference, and 4) property inference.
Each attack type exploits different aspects of the model's learned representations and training process to compromise privacy.
In the following, we will introduce these attacks and corresponding countermeasures in detail, summarized in Tabs.  \ref{tab:privacy_issues} and \ref{tab:privacy_defense}, respectively.

\begin{figure}[tbhp]
\centering
\includegraphics[width=0.9\linewidth]{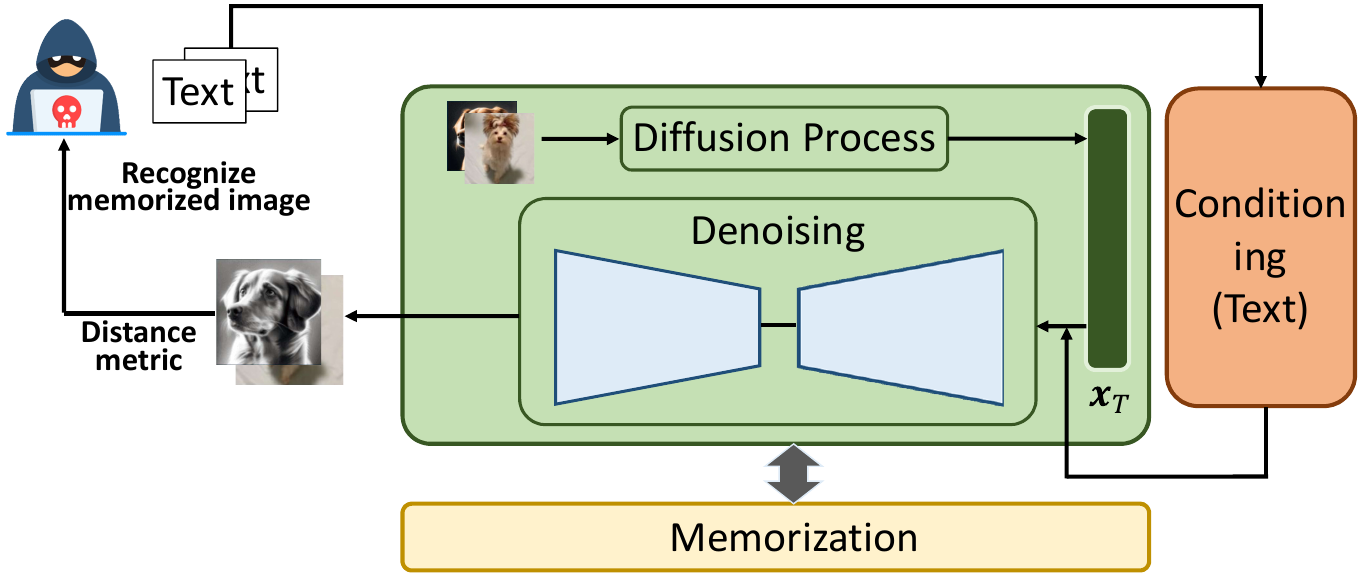}
\caption{Training data extraction in DMs refers to recovering memorized content from its generated outputs.}
\label{fig:Data extraction}
\end{figure}

\subsubsection{Training Data Extraction}
\noindent\textbf{Risks:} As shown in Fig. \ref{fig:Data extraction}, training data extraction in DMs refers to recovering, reconstructing, or identifying specific information (e.g., original training data or memorized content) from the outputs of trained DMs.
Carlini \textit{et al.}~\cite{Nicolas2023Extracting} first revealed that diffusion models are much less private than prior generative models such as GANs.
Using a generate-and-filter pipeline, Carlini \textit{et al.}~\cite{Nicolas2023Extracting} extracted over a thousand training examples from trained T2I DMs, including portraits and corporate logos.
Specifically, with Stable Diffusion, the model was queried in a black-box manner using selected prompts, generating 500 candidate images per prompt. Samples that fell within a preset similarity distance were deemed likely memorized.
Furthermore, Wen \textit{et al.}~\cite{wen2024detecting} proposed detecting memorized prompts based on the observation that memorized prompts tend to exhibit larger magnitudes of text-conditional predictions than non-memorized ones.
Specifically, it introduced a straightforward yet effective detection method centered on the magnitude of text-conditional noise predictions and then identifies memorized prompts if the detection metric falls below a tunable threshold.
This method offered precise detection without adding any extra work to the existing generation framework and even without requiring multiple generations.
Although the above training data extraction methods exhibit superior performance, their designs are limited to conditional (i.e., T2I) DMs. 
The work in \cite{chen2024extracting} proposed surrogate conditional data extraction (SIDE), a time-dependent classifier trained on generated data as surrogate conditions, which enables the extraction of training data from unconditional DMs, which are validated to be more safe in data extraction compared to conditional DMs \cite{gu2023memorization,somepalli2023understanding}.

\noindent\textbf{Countermeasures:} 
In order to prevent training data extraction,
Carlini \textit{et al.}~\cite{Nicolas2023Extracting} proposed three possible defensive methods: training data deduplication, differentially-private training, and auditing with canaries.
The experiments revealed that deduplication mitigates memorization, with a clear correlation between extraction success rates and duplication levels in large-scale models such as Stable Diffusion. 
For differentially private training, applying even non-trivial gradient clipping or noising independently (both required in DP-SGD \cite{Abadi20216Deep}) causes training to fail, highlighting the need for new advances in DP-SGD for DMs.
For auditing with canaries, we conclude that when auditing less leaky models, however, canary exposure estimates may underestimate the true data leakage.
Furthermore, Wen \textit{et al.}~\cite{wen2024detecting} proposed mitigating memorization by examining token-level changes, aiming to minimize the magnitude of text-conditional noise prediction.
Tokens undergoing large changes are considered critical for guiding predictions, while those with smaller changes are less influential.
Based on the observation that the cross-attention layers in DMs tend to overfit to embeddings of certain tokens, memorizing the corresponding training images.
Ren \textit{et al.}~\cite{Ren2024Unveiling} introduced (1) a detection approach that quantifies attention behavior patterns and (2) both inference-time and training-time strategies to reduce memorization by adjusting attention dispersion.
\subsubsection{Gradient Inversion}
\noindent\textbf{Risks:} 
Gradient inversion in DMs refers to the process of recovering or synthesizing specific data instances (e.g., training samples, sensitive inputs, or latent representations) by using the leaked gradients (typically in federated learning settings).
Unlike general data generation, reconstruction focuses on reproducing known or targeted data patterns through systematic manipulation of the diffusion process.
For instance, the work in \cite{huang2024gradient} proposed a two-phase fusion optimization, first using a well-trained generative model as prior knowledge to restrict the latent inversion search space, and then applying pixel-wise fine-tuning for refinement. This approach achieved a near-perfect reconstruction of original images while supporting privacy-preserving training, where both the Gaussian noise and sampling steps remain private.

\noindent\textbf{Countermeasures:} 
DP-SGD is the most widely used method to prevent gradient inversion attacks. However, as discussed in \cite{Nicolas2023Extracting}, the current DP-SGD method is difficult to apply for DMs, in which new advances in DP-SGD are required.

\begin{figure}[tbhp]
\centering
\includegraphics[width=1.0\linewidth]{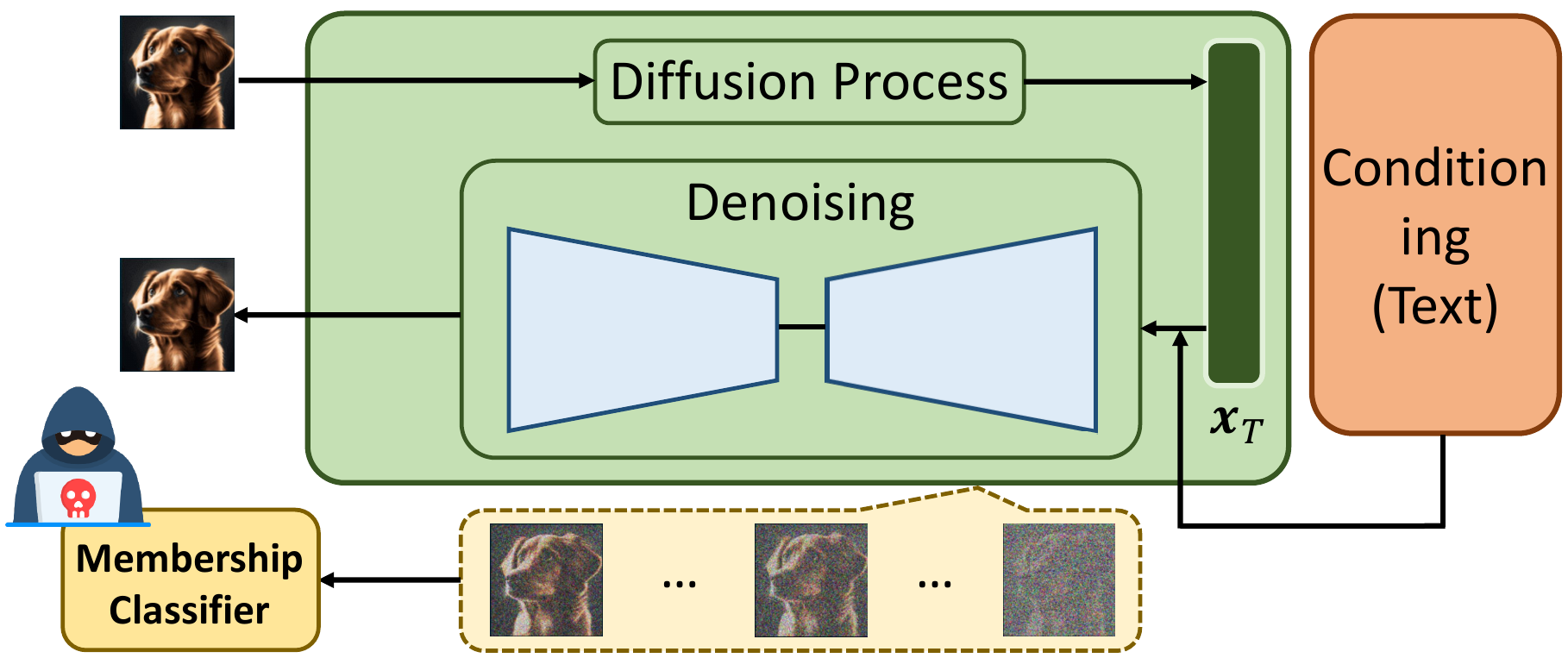}
\caption{MIA methods are techniques that determine whether a specific data sample is included in the training set of DMs.}
\label{fig:Membership inference}
\end{figure}
\subsubsection{Membership Inference}
\noindent\textbf{Risks:} 
The membership inference attack (MIA) is another classic threat in both discriminative and generative AI systems \cite{Shokri2017Membership,Chen2020GAN}, as shown in Fig. \ref{fig:Membership inference}. 
In DMs, MIA is defined to determine whether a specific sample, potentially containing sensitive or proprietary content, such as medical or personal images, was part of the model’s training set, even when the model produces only synthetic outputs.
The work in \cite{wu2022membership} presented the first privacy analysis of T2I generation models under MIA,
which designs four attack strategies based on an observation: Given a query pair (text caption $t$ and corresponding image $\boldsymbol{x}$), T2I models take $t$ as input and are optimized to generate an image $\boldsymbol{x}_{0}$ resembling the original $\boldsymbol{x}$.
Duan \textit{et al.}~\cite{Duan2023Are} then examined DM vulnerabilities that membership can be inferred by evaluating how well a query aligns with the forward-process posterior estimation at each timestep, under the assumption that members yield lower estimation errors than non-members.
Matsumoto \textit{et al.}~\cite{Matsumoto2023Membership} demonstrated that DDIM is vulnerable to overfitting attacks when trained on limited data, as it outperforms DCGAN, and that the number of sampling steps critically affects model robustness against attacks, while sampling variances show no measurable impact.
To avoid training multiple ``shadow model'', the work in ~\cite{Tang2024Membership} trained quantile regression models to estimate the distribution of reconstruction loss for non-training examples. By predicting conditional quantiles of this distribution, this attack formulated a granular hypothesis test for membership: a point is classified as a training member if its reconstruction loss falls below a dynamically computed threshold, tailored to its specific characteristics.
To reduce query budgets, the work in \cite{kong2024an} utilized the ground truth trajectory obtained by $\epsilon$ initialized at $t = 0$, and proposed an efficient query-based MIA, namely the proximal initialization attack (PIA) to infer memberships.
The work in \cite{zhai2024membership} revealed that T2I DMs exhibit stronger overfitting to the conditional distribution of images given their text prompts than to the marginal image distribution. Building on this insight, this work derived a deterministic analytical metric - Conditional Likelihood Discrepancy (CLiD) for membership inference. CLiD quantified memorization of individual samples while significantly reducing the stochasticity inherent in traditional estimation approaches.
The work in~\cite{fu2024model} introduced an MIA framework that exploits two key properties of DMs: intrinsic generative priors and Degrade Restore Compare (DRC) mechanism. This approach determines sample membership by systematically degrading the target image and comparing its restored counterpart against the original, leveraging the observation that training samples exhibit more consistent restoration behavior than non-members.
Liu \textit{et al.}~\cite{Li2024Unveiling} found that DMs better preserve structural features of training samples (members) compared to non-members during image corruption. This structural memorization effect motivated an MIA for T2I DMs, where membership is determined by measuring the structural similarity between original images and their corrupted counterparts.
Under a strict black-box assumption (i.e., remaining agnostic to model architectures and application scenarios), Zhang \textit{et al.}~\cite{Zhang2024Generated} successfully conducted MIA by only utilizing generated distributions from target generators and auxiliary non-member datasets.
The work in \cite{li2024black} conducted MIAs solely through an image-to-image variation API, requiring no access to the model's internal U-Net. This approach is motivated by the observation that DMs produce more stable noise predictions for training samples than for unseen data, determining membership by: (1) repeatedly querying the target image through the API, (2) aggregating the outputs, and (3) evaluating the consistency between the averaged predictions and the original image.
Recently, Pang \textit{et al.}~\cite{Pang2025Black} introduced the first scores-based MIA framework, grounded in a rigorous analysis of the DM objective function as its theoretical foundation, and compared the generated output against the query image to determine the membership status.

\noindent\textbf{Countermeasures:} 
DP-SGD and RandAugment are two defense methods against MIA, which cause the training of DMs to fail to converge \cite{Duan2023Are}.
In \cite{Duan2023Are}, Cutout and RandomHorizontalFlip are also two common defense methods that decrease the attack success rate (ASR) and the area under curve (AUC) of MIA to a certain degree.
Tran \textit{et al.}~\cite{tran2024dual} introduced DualMD and DistillMD to protect DMs against MIA based on training two separate DMs in disjoint subsets of the original datasets, and developed the private inference pipeline that uses both models.
Recently, the work in \cite{Luo2025Privacy} proposed SMP-LoRA, a min-max optimization framework where: (1) a proxy attack model learns to maximize membership inference gain, while (2) LDM adapts by minimizing the ratio between its adaptation loss and the attacker's membership inference gain.
\begin{figure}[t]
\centering
\includegraphics[width=1.0\linewidth]{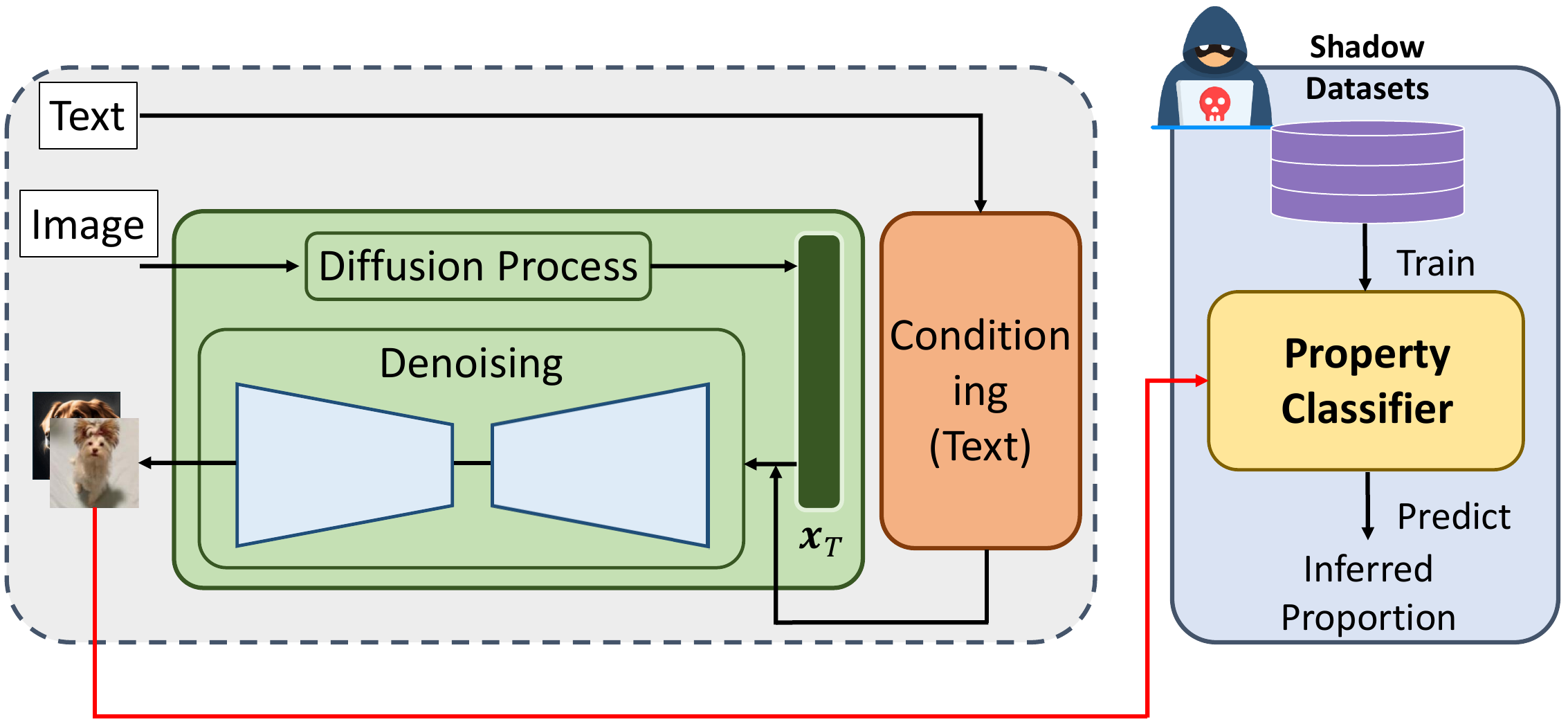}
\caption{Property inference is to extract sensitive properties of its training datasets from DMs.}
\label{fig:Property inference}
\end{figure}
\begin{table*}[h]
\caption{Taxonomy of robustness risks in DMs.}
\begin{adjustbox}{center}
\scriptsize
\begin{tabular}{|m{1.3cm}<{\centering}|m{0.3cm}<{\centering}|m{1.3cm}<{\centering}|m{2.3cm}<{\centering}|m{5.4cm}<{\centering}|m{5.0cm}<{\centering}|}
\hline
\textbf{Category}&\textbf{Ref.}& \textbf{Attacker's knowledge}& \textbf{Target Models}& \textbf{Effectiveness}&\textbf{Limitations}\\
\hline\hline
\multirow{22}*{\shortstack{Adversarial\\ attacks}}
&\cite{Zhuang2023A}&Gray-box& SD&Enable precise directional DM editing, erasing targets while preserving other regions with minimal distortion
&Access to the CLIP text encoder\\
\cline{2-6}
&\cite{Haz2024asymmetric}&White-box& SD v2-1&Reveal ASR differences when bidirectionally swapping entities, showing an asymmetry in adversarial attacks&Results show attack success probabilities spanning 60\% to below 5\%, depending on defense configurations\\
\cline{2-6}
&\cite{shahariar2024adversarial}&Black-box&SD v1.5&Critical tokens and content fusion are part-of-speech (POS)-dependent, whereas suffix transferability is category-invariant& The metrics utilized in this study to assess the attack may not fully capture the visual plausibility or semantic accuracy of images after the attack\\
\cline{2-6}
&\cite{Li2024VA3}& Black-box & SD-v1-4&The probability of generating infringing content grows monotonically with interaction length, with each query having a minimum success probability &Development of more robust copyright protection approaches\\
\cline{2-6}
&\cite{Yu2024Step}&Black-box & LDM & The
reverse process of the DMs is susceptible to the shift of the mean noise value & Choosing vulnerable steps to attack can further improve the attacking performance is not clear\\
\cline{2-6}
&\cite{liu2024discovering}& Black-box &SD, DeepFloyd, GLIDE&Investigate discrete prompt space and high-dimensional latent space to automatically identify failure modes in image generations& Rough surrogate loss functions and vanishing gradient problem\\
\cline{2-6}
&\cite{zeng2024advi2i}& White-box & Instruct-Pix2Pix, SDv1.5& Dynamically adapt to countermeasures by reducing similarity between adversarial images and NSFW embeddings in latent space, enhancing robustness &Investigate robust defenses against adversarial image attacks, considering
both text and image inputs when designing safety mechanisms for generative models\\
\cline{2-6}
&\cite{Zhang2025To}&Black-box&SD v1.4&Leverage DMs' inherent classification to generate adversarial prompts without extra classifiers or DMs&Only evaluation on SDv1.4\\
\hline
\end{tabular}
\end{adjustbox}
\label{tab:robustness_issues}
\end{table*}
\begin{table*}[t]
\caption{Taxonomy of defense methods against robustness issues in DMs.}
\centering
\scriptsize
\begin{tabular}{|m{2cm}<{\centering}|m{0.6cm}<{\centering}|m{3.5cm}<{\centering}|m{2.5cm}<{\centering}|m{7.4cm}<{\centering}|}
\hline
\textbf{Threats}& \textbf{Ref.}&\textbf{Key Methods}& \textbf{Target Models}& \textbf{Effectiveness}\\
\hline\hline
\multirow{5}*{\shortstack{Adversarial\\ attacks}}&\multirow{1}*{\shortstack{\cite{Liu2025Latent}}}&Latent Guard&SD v1.5, SDXL& Allow for test time modifications of the blacklist, without retraining needs\\
\cline{2-5}
&\multirow{1}*{\shortstack{\cite{Zhang2025ProTIP}}}&ProTIP&SD-V1.5, SD-V1.4, SDXL-Turbo& Incorporate several sequential analysis methods to dynamically determine the sample size and thus enhance the efficiency\\
\cline{2-5}
&\cite{qiu2025safe} & Embedding Sanitizer &SD-v1.4&Not only mitigates the generation of harmful concepts but also improves interpretability and controllability\\
\hline
\end{tabular}
\label{tab:robustness_defense}
\end{table*}
\begin{table*}[h]
\caption{Taxonomy of safety issues in DMs.}
\begin{adjustbox}{center}
\scriptsize
\begin{tabular}{|m{1.3cm}<{\centering}|m{0.3cm}<{\centering}|m{1.3cm}<{\centering}|m{2.3cm}<{\centering}|m{6.5cm}<{\centering}|m{5.0cm}<{\centering}|}
\hline
 \textbf{Category}& \textbf{Ref.}& \textbf{Attacker's knowledge}& \textbf{Target Models}& \textbf{Effectiveness}&\textbf{Limitations}\\
\hline\hline
\multirow{10}*{\shortstack{Backdoor}}
&\cite{Chou2023How}&White-Box&DDPM&Generate the target image once the initial noise or the initial image contains the backdoor trigger& Only consider DDPM \\
\cline{2-6}
&\cite{Chen2023TrojDiff}&White-Box&DDPM,  DDIM&Reveal DM vulnerabilities to training data manipulations by backdooring with novel transitions that diffuse adversarial targets into a biased Gaussian distribution& 1) Only consider DDPM and DDIM; 2) Induce a higher FID than benign models\\
\cline{2-6}
&\cite{Zhai2023Text}&White-Box& SD v1.4& Inject various backdoors into the model to achieve different pre-set goals& Without considering advanced defense methods\\
\cline{2-6}
&\cite{Chou2023VillanDiffusion}&White-Box& DDPM, LDM, NCSN&The generality of our unified backdoor attack on a variety of choices in DMs, samplers, and unconditional/conditional generations&Without covering all kinds of DMs, like Cold Diffusion and Soft Diffusion\\
\cline{2-6}
&\cite{zhang2025invisible}&White-Box& SD v1.4&Achieve a 97.5\% attack success rate while exhibiting stronger resistance to defenses, with over 98\% backdoor samples bypassing three SOTA detection mechanisms&Bring extra time consumption compared to previous works\\
\cline{1-6}
\multirow{22}*{\shortstack{Jailbreak}}&\cite{ma2024jailbreak}& Black-box &Open sourced T2I, MidJourney& Bypass both text and image safety checkers&Concept pairs are given by ChatGPT, which needs prompting that out of the automated framework\\
\cline{2-6}
&\cite{Qu2023Unsafe}&White-box&SD, LDM, DALL$\cdot$E 2-demo, DALL$\cdot$E mini& Easily generate realistic hateful meme variants&Full access to the T2I model, i.e., the adversary can modify model parameters to personalize image generation\\
\cline{2-6}
&\cite{Yang2024MMA}&White-Box, Black-Box&SD, Midjounery, Leonardo.Ai& Bypass prompt filters and safety checkers& Ideal prompt filters and safety checkers\\
\cline{2-6}
&\cite{Huang2025Perception}&Black-box&DALL·E 2, DALL·E 3, Cogview3, SDXL, Tongyiwanxiang, Hunyuan&Require no specific T2I architecture, and produce highly natural adversarial prompts that maintain semantic coherence& Lack classic models, such as Midjounery and
Leonardo.Ai\\
\cline{2-6}
&\cite{Ma2024ColJailBreak}&Black-box&GPT-4, DALL·E 2&Adaptive normal safe substitution, inpainting-driven injection of unsafe content, and contrastive language-image-guided collaborative optimization&Complex prompts or multiple sensitive words require repeated edits, reducing efficiency and coherence of generated content\\
\cline{2-6}
&\cite{gao2024htsattack}&Black-box&SDv1.4, SafeGen, SLD&Efficiently bypass the latest defense mechanisms, including prompt, posthoc image checkers, secure T2I, and online commercial models&Lack commercial models, such as DALL·E 3, Midjounery, and
Leonardo.Ai\\
\cline{2-6}
&\cite{zhao2025inception}&Black-box&DALL·E 3&Recursively split unsafe words into benign chunks, ensuring no semantic loss while bypassing safety filters& Consider DALL·E 3, without Midjounery and
Leonardo.Ai\\
\cline{2-6}
&\cite{liu2025token}&Black-box&SDv1.4, SLD (Medium), SafeGen, DALL-E 3&Produce semantic adversarial prompts that successfully evade multi-layered defense mechanisms in T2I models&Consider DALL·E 3, without Midjounery and
Leonardo.Ai\\
\cline{2-6}
&\cite{Yang2024Sneaky}&Black-box&DALL·E 2&Iteratively query T2I, perturbing prompts via feedback to bypass safety filters while maintaining quality&Consider DALL·E 2, without other powerful commercial models \\
\hline
\end{tabular}
\label{tab:safety_issues}
\end{adjustbox}
\end{table*}
\begin{table*}[h]
\caption{Taxonomy of defense methods against safety issues in DMs.}
\centering
\scriptsize
\begin{tabular}{|m{1.2cm}<{\centering}|m{0.3cm}<{\centering}|m{3.7cm}<{\centering}|m{2.5cm}<{\centering}|m{8.3cm}<{\centering}|}
\hline
\textbf{Threats}& \textbf{Ref.}&\textbf{Key Methods}& \textbf{Target Models}& \textbf{Effectiveness}\\
\hline\hline
\multirow{5}*{\shortstack{Backdoor}}&\multirow{1}*{\shortstack{\cite{Wang2025T2IShield}}}&T2IShield&SD v1.4&Achieve an 88.9\% detection F1 score with minimal computational overhead, along with an 86.4\% localization F1 score, while successfully invalidates 99\% of poisoned samples
\\
\cline{2-5}
&\multirow{1}*{\shortstack{\cite{yoon2025safree}}}&SAFREE&SD v1.4&Maintain consistent safety verification while preserving output fidelity, quality, and safety\\
\cline{2-5}
&\cite{Guan2025UFID} & UFID & SD v1.4& Achieve exceptional detection accuracy while maintaining superb computational efficiency\\
\cline{1-5}
\multirow{4}*{\shortstack{Jailbreak}} &\multirow{1}*{\shortstack{\cite{tian2025sparse}}}
&Interpret then Deactivate&SD1.4&Robustly erase target concepts without forgetting on remaining concepts\\
\cline{2-5}
&\multirow{1}*{\shortstack{\cite{liu2024safetydpo}}}&SafetyDPO&SD v1.5, SDXL&Enable T2I models to generate outputs that are not only of high quality but also aligned with safety and ethical guidelines\\
\cline{2-5}
&\multirow{1}*{\shortstack{\cite{yoon2025safree}}}&SAFREE &SD-v1.4&Show competitive results against training-based methods\\
\hline
\end{tabular}
\label{tab:safety_defense}
\end{table*}
\subsubsection{Property Inference}
\noindent\textbf{Risks:} As shown in Fig. \ref{fig:Property inference}, different from MIA, the goal of property inference attacks is to extract sensitive global properties of its training datasets from an ML model~\cite{zhou2021property}, i.e., DM in this paper.
Hu \textit{et al.} \cite{hu2024prisampler} first conducted a comprehensive evaluation of property inference attacks on various DMs trained on diverse data types, including tabular and image datasets, and found that DMs and their samplers are universally vulnerable to property inference attacks through a broad range of evaluations.

\noindent\textbf{Countermeasures:} 
The proposed PriSampler \cite{hu2024prisampler} is a plug-in model-agnostic defense mechanism designed to mitigate property inference risks in DMs by strategically manipulating the sampling process to obscure sensitive property distributions. This method operates by: (1) identifying property decision hyperplanes in the DM's latent space, and (2) guiding the sampler to generate samples that strategically perturb these property correlations while maintaining output quality.
\subsection{Robustness}\label{subsec:robustness}
Adversarial attacks pose a significant threat to the robustness of DMs to manipulate the generated content by exploiting input perturbations. Here, we mainly investigate the vulnerabilities to adversarial attacks, summarized in Tabs. \ref{tab:robustness_issues} and \ref{tab:robustness_defense}.
\begin{figure}[tbhp]
\centering
\includegraphics[width=1.0\linewidth]{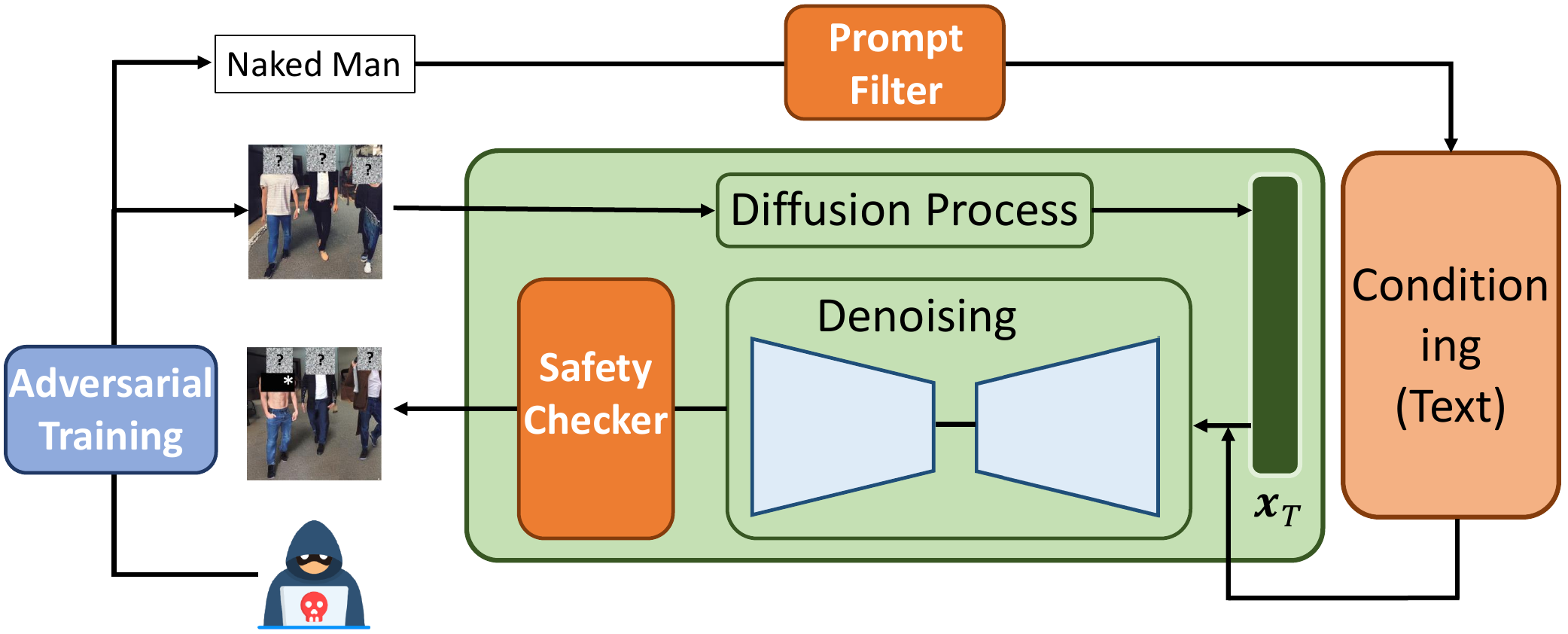}
\caption{Adversarial attacks inject precisely crafted perturbations into inputs to induce deviations in DMs' outputs.}
\label{fig:Adversarial attacks}
\end{figure}

\noindent\textbf{Risks:} Adversarial attacks against ML models have attracted significant attention in recent years due to their evident harmfulness, such as misleading signs in autonomous driving, as shown in Fig. \ref{fig:Adversarial attacks}.
Extending this concern to DMs, several studies have demonstrated their susceptibility to adversarial attacks, where adversaries aims to manipulate the model into generating malicious content or incorrect outputs by carefully crafting perturbations to inputs (such as text prompts or initial noise).
Zhuang \textit{et al.}~\cite{Zhuang2023A} introduced both untargeted and targeted adversarial attacks in the absence of end-to-end model queries, which exploit the most influential dimensions in the text embedding space, termed steerable key dimensions. 
Their results demonstrate that a mere five-character modification in the input prompt induces substantial shifts in the synthesized images from Stable Diffusion.
Building on this, Shahgir \textit{et al.} \cite{Haz2024asymmetric} proposed entity swapping as a new attack objective using adversarial suffixes, supported by two gradient-based attack methods.
Similarly, Shahariar \textit{et al.} \cite{shahariar2024adversarial} constructed a high-quality dataset for realistic POS tag swapping and employed gradient-based attacks to discover adversarial suffixes that fool T2I models into generating images with swapped tokens.

Furthermore, Yu \textit{et al.}~\cite{Yu2024Step} revealed that susceptibility to adversarial attacks varies significantly between different reverse steps. By modeling step-wise vulnerability and using it to guide step selection, their approach substantially improved adversarial effectiveness.  
Liu \textit{et al.} \cite{liu2024discovering} proposed SAGE, the first adversarial search method to systematically explore both discrete prompt space and high-dimensional latent space, to uncover generation failures and undesirable behaviors.
To generate Not Safe for Work (NSFW) content, Zeng \textit{et al.}~\cite{zeng2024advi2i} introduced AdvI2I, a framework that optimizes an image generator to craft adversarial outputs capable of bypassing defenses such as Safe Latent Diffusion (SLD), while preserving original text prompts.
To further adapt to potential countermeasures, they developed AdvI2I-Adaptive, which minimizes similarity to NSFW embeddings and enhances attack resilience.\\
\noindent\textbf{Countermeasures:} Several countermeasures have been proposed to mitigate these vulnerabilities. 
Liu \textit{et al.}~\cite{Liu2025Latent} proposed detecting blacklisted concepts in a learned latent space above the text encoder, 
enabling detection beyond literal word matching. Their approach allows for dynamic modification of the blacklist during inference, supporting concept addition or removal without retraining the model.  
Zhang \textit{et al.} \cite{Zhang2025ProTIP} employed sequential analysis with efficacy/utility stopping rules for adversarial detection, combined with adaptive concentration inequalities to dynamically optimize the number of stochastic perturbations when verification criteria are satisfied.
Similarly, Li \textit{et al.}~\cite{Li2024SafeGen} presented a text-agnostic framework that prevents the generation of sexual content in T2I models by internally suppressing explicit visual representations, making DMs inherently resistant to generating NSFW content.
From the perspective of unlearning, Park \textit{et al.} \cite{park2024direct} proposed Direct Unlearning Optimization (DUO), a framework that selectively removes NSFW content from T2I models while maintaining performance on benign concepts. DUO leveraged preference optimization with curated image pairs to unlearn unsafe visual representations without compromising unrelated features.
Recently, Hu \textit{et al.}~\cite{hu2025safetext} proposed SafeText, an alignment approach that selectively fine-tunes the text encoder, rather than the DM, to substantially modify the embeddings of unsafe prompts while preserving those of safe prompts.
\begin{table*}[t]
\centering
\scriptsize
\caption{Taxonomy of copyright issues in DMs.}
\begin{tabular}{|m{1.3cm}<{\centering}|m{0.3cm}<{\centering}|m{1.3cm}<{\centering}|m{2.3cm}<{\centering}|m{5.4cm}<{\centering}|m{5.0cm}<{\centering}|}
\hline
\textbf{Category}&\textbf{Ref.}& \textbf{Attacker's knowledge}& \textbf{Target Models}& \textbf{Effectiveness}&\textbf{Limitations}\\
\hline\hline
Model extraction&\cite{Mahajan2024Prompting} &White-box&SD v1.5, LDM&Yield meaningful prompts that synthesize accurate, diverse images of a target concept& Without considering commercial models like DALL·E 3,
Midjourney and Leonardo.Ai\\
\hline
Prompt stealing&
\cite{Shen2024Prompt}&Black-box&SD (v1.4, v1.5, v2.0), Midjourney, DALL$\cdot$E 2&Identify the modifiers within the generated image&Require a strong assumption for the defender, i.e., white-box access to the attack model \\
\hline
Data misuse&-&-&-&-&- \\
\hline
\end{tabular}
\label{tab:copyright_issues}
\end{table*}
\begin{table*}[h]
\caption{Taxonomy of defense methods against copyright issues in DMs.}
\centering
\scriptsize
\begin{tabular}{|m{1.2cm}<{\centering}|m{0.3cm}<{\centering}|m{3.7cm}<{\centering}|m{2.5cm}<{\centering}|m{8.3cm}<{\centering}|}
\hline
\textbf{Threats}& \textbf{Ref.}&\textbf{Key Methods}& \textbf{Target Models}& \textbf{Effectiveness}\\
\hline\hline
Model extraction&\multirow{1}*{\shortstack{\cite{liu2023watermarkingdiffusionmodel}}}&NAIVEWM, FIXEDWM&LDM& Inject the watermark into the DMs and can be verified by the pre-defined prompts\\
\hline
Prompt stealing&\multirow{1}*{\shortstack{\cite{Yao2024PromptCARE}}}&PromptCARE&BERT, RoBERTa, Facebook OPT-1.3b& Watermark and verification schemes specifically designed for unique properties of prompts and the natural language domain\\
\hline
\multirow{5}*{\shortstack{Data\\misuse}}&\multirow{1}*{\shortstack{\cite{Zhu2025TabWak}}}&TabWak&Tabular DM& Control the sampling of Gaussian latents for table row synthesis through the diffusion backbone\\
\cline{2-5}
&\multirow{1}*{\shortstack{\cite{Cheng2023Flexible}}}&End-to-end watermarking&LDM&Enable the embedded message in the generated image to be modified as needed without retraining or fine-tuning the LDM\\
\cline{2-5}
&\multirow{1}*{\shortstack{\cite{Cui2025FT-Shield}}}&FT-Shield&LDM&Watermark transfer from source images to generated content, enabling copyright verification\\
\hline
\end{tabular}
\label{tab:copyright_defense}
\end{table*}
\begin{figure}[tbhp]
\centering
\includegraphics[width=1.0\linewidth]{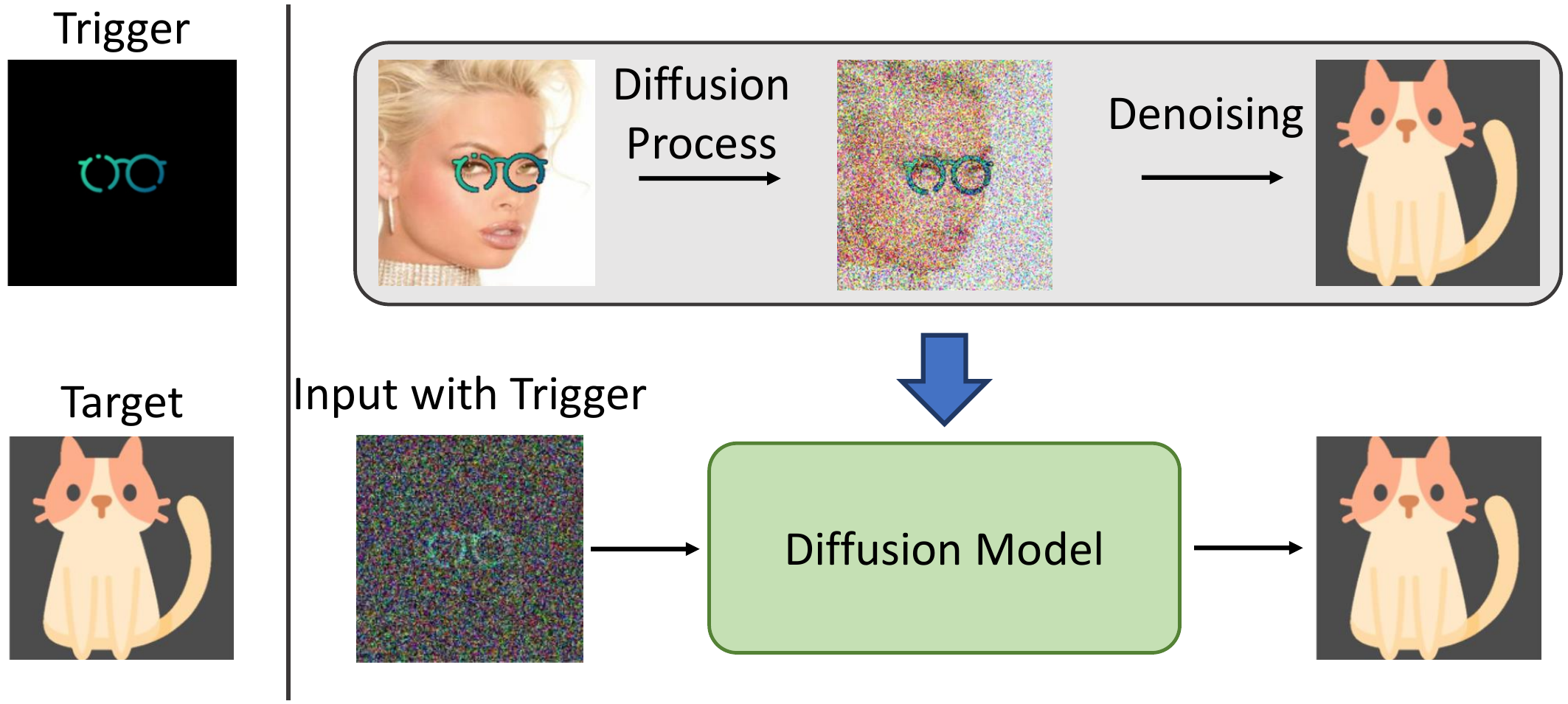}
\caption{The backdoor attack alters the DMs' forward process, mapping the backdoor target's distribution to a poisoned image with standard Gaussian noise.}
\label{fig:Backdoor attack}
\end{figure}
\subsection{Safety}\label{subsec:safety}
Safety concerns in DMs are predominantly centered around backdoor and jailbreak attacks, summarized in Tabs. \ref{tab:safety_issues} and \ref{tab:safety_defense}. We survey these two attack categories below.

\subsubsection{Backdoor}
\noindent\textbf{Risks:}
The first safety issue is the backdoor attack, as shown in Fig. \ref{fig:Backdoor attack}.
When hidden backdoors are activated by attacker-specified triggers, they maliciously alter ML model outputs, making this a critical research focus \cite{Li2024Backdoor}.
Prior work has revealed that both unconditional and conditional DMs are vulnerable to backdoor injection \cite{Chou2023How,Chen2023TrojDiff,Zhai2023Text,Chou2023VillanDiffusion}.
For unconditional DMs, Chou \textit{et al.}~\cite{Chou2023How} modified the forward process of DDPM into a backdoor variant mapping the target backdoor distribution to poisoned images with standard Gaussian noise.
Similarly, Chen \textit{et al.}~\cite{Chen2023TrojDiff} assumed that an attacker with access to both training and sampling mechanisms can inject correction terms into DDPM and DDIM to execute the attack.
For conditional DMs, Zhai \textit{et al.}~\cite{Zhai2023Text} implemented backdoor attacks through (1) a regularization loss to avoid overfitting on target patches and (2) a teacher model to preserve normal functionality with benign inputs.
Extending this, Chou \textit{et al.}~\cite{Chou2023VillanDiffusion} proposed a unified backdoor attack framework that formulates backdooring unconditional DMs as a distribution mapping problem and generalizes it to conditional generation by adding clean and backdoored captions.
Recently, Zhang \textit{et al.} \cite{zhang2025invisible} introduced an Invisible Backdoor Attack (IBA) explicitly targeting two key consistencies in T2I DMs. First, IBA exploited syntactic structures as triggers to amplify sensitivity to textual variations, effectively breaking semantic consistency. Second, IBA applied a Kernel Maximum Mean Discrepancy-based regularization to align cross-attention responses between backdoored and benign samples, thereby disrupting attention consistency. Together, these mechanisms enabled a stealthy yet potent attack while maintaining model utility on benign inputs.

\noindent\textbf{Countermeasures:}
Although backdoor detection and removal methods have been widely studied in traditional ML systems, they cannot be applied to DMs directly, as the iterative generative process, high-dimensional latent space, and T2I conditioning introduce fundamentally different attack surfaces and behaviors.
To address this, Wang \textit{et al.}~\cite{Wang2025T2IShield} proposed T2IShield, which is based on ``Assimilation Phenomenon" on the
cross-attention maps caused by the backdoor trigger, and introduces two effective backdoor detection methods.
T2IShield further introduced a binary-search-based method for precise trigger localization and evaluated the use of concept-editing techniques as defensive measures.  
SAFREE was proposed in \cite{yoon2025safree}, a training-free plug-and-play mechanism designed to ensure safe generation without modifying pre-trained weights. By filtering unsafe concepts in both the textual prompt embedding space and visual latent space, SAFREE preserves fidelity, quality, and efficiency while enhancing robustness.  
From the perspective of causal analysis, Guan \textit{et al.}~\cite{Guan2025UFID} grounded the problem in a causal analysis and revealed that backdoor attacks act as confounders, creating spurious associations between input prompts and target outputs. Critically, this confounding effect persisted even under Gaussian noise perturbations, highlighting the need for causal inference-based defenses.  

\subsubsection{Jailbreak}
\noindent\textbf{Risk:} Unlike adversarial attacks that primarily manipulate outputs using imperceptible inputs, jailbreak attack aims to generate inappropriate or unsafe context with/without adversarial attacks to break through the model's usage restrictions and ethical constraints~\cite{Qu2023Unsafe,Yang2024MMA,ma2024jailbreak}. 
Qu \textit{et al.} \cite{Qu2023Unsafe} first revealed that the vulnerabilities of the T2I model lie in the inherent weaknesses of text encoder robustness.
Furthermore, Yang \textit{et al.}~\cite{Yang2024MMA} introduced MMA-Diffusion, which exploits multimodal (textual and visual) vulnerabilities to bypass both prompt filters and post-generation safety checkers, exposing fundamental flaws in current defense mechanisms.
To make jailbreak attacks more systematic, Ma \textit{et al.} \cite{ma2024jailbreak} proposed a two-stage approach: (1) searching the text embedding space using ChatGPT-generated antonym groups, followed by (2) discrete vocabulary optimization of prefix prompts to semantically align target concepts.
Similarly, Gao \textit{et al.}~\cite{gao2024htsattack} proposed HTS-Attack, which removes sensitive tokens before performing heuristic search through recombination and mutation of high-performing candidates. By maintaining population diversity with both optimal and suboptimal prompts, HTS-Attack effectively bypasses defenses while avoiding local optima.

Other works focused on query efficiency and the naturalness of jailbreak prompts. 
Liu \textit{et al.}~\cite{liu2025token} proposed TCBS-Attack, a black-box method that identifies and optimizes boundary-proximate tokens to craft semantically-preserving adversarial prompts capable of evading multilayer defenses.
Huang \textit{et al.} \cite{Huang2025Perception} proposed perception-guided jailbreak (PGJ), an LLM-driven approach that generates natural adversarial prompts by replacing unsafe words with perceptually similar but semantically inconsistent safe phrases, requiring no model-specific knowledge (model-free).
In addition, Ma \textit{et al.} \cite{Ma2024ColJailBreak} presented ColJailBreak, which combines three components: adaptive normal safe substitution, inpainting-driven injection of unsafe content, and collaborative optimization guided by contrast images.
Yang \textit{et al.}~\cite{Yang2024Sneaky} introduced SneakyPrompt, which iteratively queries the T2I model and perturbs prompt tokens based on feedback to evade safety filters.

\noindent\textbf{Countermeasures:} 
To mitigate jailbreak risks, Tian \textit{et al.}~\cite{tian2025sparse} repurposed SAE as zero-shot classifiers that detect whether input prompts include target unsafe concepts. By permanently deactivating specific features linked to harmful concepts, their approach enables selective concept erasure. 
Liu \textit{et al.}~\cite{liu2024safetydpo} presented SafetyDPO, which aligns T2I models with safety objectives through Direct Preference Optimization (DPO). To enable DPO, they introduced CoProV2, a synthetic dataset of harmful and safe image-text pairs, and a merging strategy to consolidate LoRA experts into a single model to achieve scalable performance.   
\begin{table*}[h]
\centering
\scriptsize
\caption{Taxonomy of fairness issues in DMs.}
\begin{tabular}{|m{1.3cm}<{\centering}|m{0.3cm}<{\centering}|m{1.3cm}<{\centering}|m{2.3cm}<{\centering}|m{5.4cm}<{\centering}|m{5.0cm}<{\centering}|}
\hline
\textbf{Category}&\textbf{Ref.}& \textbf{Attacker's knowledge}& \textbf{Target Models}& \textbf{Effectiveness}&\textbf{Limitations}\\
\hline\hline
\multirow{1}*{\shortstack{Fairness}}&\cite{Luo2024Exploring}&Black-box&NCSN, DDPM, SDEM, TabDDPM&Fairness poisoning attacks: manipulate training data distribution to compromise the integrity of downstream models&Limited generalization to other models like GANs and VAEs\\
\hline
\end{tabular}
\label{tab:fairness_issues}
\end{table*}
\begin{table*}[h]
\centering
\scriptsize
\caption{Taxonomy of defense methods against fairness issues in DMs.}\label{tab:attacks}
\begin{tabular}{|m{1.2cm}<{\centering}|m{0.5cm}<{\centering}|m{3.7cm}<{\centering}|m{2.5cm}<{\centering}|m{8.1cm}<{\centering}|}
\hline
\textbf{Threats}& \textbf{Ref.}&\textbf{Key Methods}& \textbf{Target Models}& \textbf{Effectiveness}\\
\hline\hline
\multirow{7}*{\shortstack{Fairness}}
&\multirow{1}*{\shortstack{\cite{friedrich2023fair}}}&Fair Diffusion&SD v1.5& Use learned biases and user guidance to steer the model toward a specified fairness goal during inference\\
\cline{2-5}
&\multirow{1}*{\shortstack{\cite{shen2024finetuning}}}&DFT, Adjusted DFT&Runwayml/SD-v1-5 & An alignment loss that steers image generation toward target distributions using adjusted gradients to optimize on-output losses\\
\cline{2-5}
&\multirow{1}*{\shortstack{\cite{Jiang2024DifFaiRec}}}&DifFaiRec&Diffusion Model& Design a counterfactual module to reduce the model sensitivity to protected attributes and provide mathematical explanations\\
\cline{2-5}
&\multirow{1}*{\shortstack{\cite{huang2025debiasing}}}&DDM&SD v1.5, SD v2& Learn debiased latent representations via indicator-guided training, ensuring fairness without predefined sensitive attributes\\
\hline
\end{tabular}
\label{tab:fairness_defense}
\end{table*}
\subsection{Copyright}\label{subsec:copyright}
Copyright aims to protect original human-authored works. As DMs have been widely used in visual art, such as art design~\cite{wang2025diffusion}, copyright issues have drawn a lot of attention in DMs \cite{Vyas2022On}.
Copyright issues in DMs mainly represent model extraction, prompt stealing, and data misuse.
We summarize copyright issues and countermeasures in Tabs. \ref{tab:copyright_issues} and \ref{tab:copyright_defense}, respectively.
\subsubsection{Model Extraction}
\noindent\textbf{Risk:} The model extraction attack was introduced by Tram{\`{e}}r~\cite{Florian2016Stealing}, in which adversaries exploit query interfaces through malicious attacks to extract and steal proprietary model parameters. 
More precisely, adversaries leveraged systematic query interactions to progressively learn and replicate sensitive server-side models through black-box access only. 
To date, there is a lack of research on model extraction or stealing targeting DMs.
This presents a promising direction, i.e., how a well-trained DM can be replicated or approximated using only limited access or a small amount of training data.

\noindent\textbf{Countermeasures:} 
In order to defend model extraction, the work in \cite{liu2023watermarkingdiffusionmodel} proposed two effective watermarking schemes, i.e., NAIVEWM and FIXEDWM, to protect the intellectual property by injecting the watermark into the DMs, and can be verified by pre-defined prompts or prompts containing a trigger at a fixed position.
\subsubsection{Prompt Stealing}
\noindent\textbf{Risk:} Shen \textit{et al.} \cite{Shen2024Prompt} first introduced the novel threat of prompt stealing attacks, where adversaries extract text prompts from images generated by T2I models. This work presented PromptStealer, a two-stage framework comprising: (1) a subject generator that reconstructs the core subject from generated images, and (2) a modifier detector that identifies stylistic and compositional modifiers.
Furthermore, Mahajan \textit{et al.}~\cite{Mahajan2024Prompting} proposed an inversion approach against DMs for direct, interpretable prompt recovery using a delayed projection scheme that optimizes prompts within the model's vocabulary space. This method is built on the key insight that different diffusion timesteps capture distinct hierarchical features, with later noisy timesteps encoding semantic information and recovered text tokens that accurately represent the image's core meaning.
The work in \cite{Li2024VA3} introduced Virtually Assured Amplification Attack (VA3), with an amplification effect where sustained interactions increase the likelihood of generating infringing content, and (2) a provable non-trivial lower bound on the success probability for each model engagement.
Recently, Zhao \textit{et al.}~\cite{zhao2025towards} revealed the poor adaptability of existing prompt stealing attacks and proposed effective training-free method that generates modifiers on
the fly and leverages feedbacks from proxy to optimize
the prompt effectively.

\noindent\textbf{Countermeasures:} 
The risks of prompt stealing are studied from various aspects, but the schemes for prompt copyright protection are unexplored.
For LLMs, Yao \textit{et al.} \cite{Yao2024PromptCARE} proposed PromptCARE, watermark injection, and verification schemes, which are customized to the characteristics pertinent to prompts and the domain of natural languages.

\subsubsection{Data Misuse}
\noindent\textbf{Risk:} The data misuse refers to any illegal or irresponsible handling of data in DMs. We can divide data misuse into three categories: unauthorized utilization of data during the training or fine-tuning, direct reproduction of training datasets, and improper use of generated outputs.

\noindent\textbf{Countermeasures:} 
In order to prevent unauthorized utilization of data, Wang \textit{et al.} \cite{Wang2024DIAGNOSIS} proposed a method for detecting such unauthorized data usage by planting the injected memorization into the T2I DMs trained on the protected dataset.
For the direct reproduction of training dataset, Somepalli \textit{et al.} \cite{Somepalli2023Diffusion} proposed an image retrieval framework that compares generated images with training samples and detects when content has been replicated. 
For improper use of generated outputs, the work in \cite{Cui2025FT-Shield} proposed FT-Shield, a watermarking solution, addressing copyright protection challenges by developing novel watermark generation and detection mechanisms that guarantee robust watermark transfer from training data to model outputs, enabling reliable identification of copyrighted content utilization.
Cheng \textit{et al.}~\cite{Cheng2023Flexible} proposed an end-to-end watermarking system that: (1) supports adaptable message embedding without requiring model fine-tuning, and (2) systematically prevents users from generating non-watermarked images.
Furthermore, TabWak \cite{Zhu2025TabWak} proposed a row-wise watermarking framework for tabular DMs that embeds imperceptible patterns while preserving data utility and maintaining robust detectability against post-generation modifications.
\subsection{Fairness}\label{subsec:fairness}
Fairness is a classic research topic in generative models, such as fair GANs \cite{Christopher2023On}, which is also existing in DMs. We summarize the fairness issues and countermeasures of DMs in Tabs. \ref{tab:fairness_issues} and \ref{tab:fairness_defense}.

\noindent\textbf{Risks:} 
DMs learn to generate data by approximating the probability distribution of their training datasets. 
If these datasets contain biases, DMs may reflect and often amplify these biases.
Therefore, there has been an increase in interest in fair DMs, such as \cite{Luo2024Exploring}.
In \cite{Luo2024Exploring}, the authors investigated the potential risks of privacy and fairness associated with the sharing of DMs via manipulating the distribution of the training dataset of DMs.

\noindent\textbf{Countermeasures:} 
The first practical framework to enforce fairness in DMs was presented in \cite{friedrich2023fair}, identifying and mitigating biased concepts learned during training by adaptively steering model generations toward fairer outputs during inference. Crucially, the system maintains user control, allowing human-guided fairness adjustments through natural language instructions.
Shen \textit{et al.} \cite{shen2024finetuning} formulated fairness as a distributional alignment challenge, addressed through two key innovations: (1) a distributional alignment loss that guides generated image characteristics toward user-specified target distributions; (2) Adjusted Direct Fine-Tuning (DFT), which modifies the sampling process via gradient adjustments to directly optimize image-space loss functions.
Jiang \textit{et al.} \cite{Jiang2024DifFaiRec} proposed DifFaiRec to provide fair recommendations built upon the conditional DMs. To guarantee fairness, this work designed a counterfactual module to reduce the model sensitivity to protected attributes.
The work in \cite{huang2025debiasing} leveraged an indicator to decouple sensitive attributes from learned representations, advancing fairness by learning balanced representations, eliminating the need for explicit sensitive attribute specification.

\begin{table*}[h]
\caption{Taxonomy of hallucination issues in DMs.}
\centering
\scriptsize
\begin{tabular}{|m{1.2cm}<{\centering}|m{0.5cm}<{\centering}|m{1.3cm}<{\centering}|m{2.3cm}<{\centering}|m{5.9cm}<{\centering}|m{4.3cm}<{\centering}|}
\hline
\textbf{Category}&\textbf{Ref.}& \textbf{Attacker's knowledge}& \textbf{Target Models}& \textbf{Effectiveness}&\textbf{Limitations}\\
\hline\hline
\multirow{3}*{\shortstack{Hallucina-\\tion}}
&\cite{qin2024evaluating}&Black-box&SD v1-4, SD v2, SD XL&Provide human-aligned, intuitive comprehensive scoring& Struggle to effectively detect key objects in synthesized landscape images\\
\cline{2-6}
&\cite{Aithal2024Understanding}&Black-box&DDPM&Hallucination is indicated by high variance in the sample’s trajectory during the final backward steps& The selection of the right timesteps is key to detect hallucinations\\
\hline
\end{tabular}
\label{tab:hallucination_issues}
\end{table*}
\begin{table*}[h]
\caption{Taxonomy of defense methods against hallucination issues in DMs.}
\centering
\scriptsize
\begin{tabular}{|m{1.2cm}<{\centering}|m{0.5cm}<{\centering}|m{3.7cm}<{\centering}|m{2.5cm}<{\centering}|m{8.1cm}<{\centering}|}
\hline
\textbf{Threats}& \textbf{Ref.}&\textbf{Key Methods}& \textbf{Target Models}& \textbf{Effectiveness}\\
\hline\hline
\multirow{6}*{\shortstack{Hallucina-\\tion}}
&\multirow{1}*{\shortstack{\cite{Kim2025Tackling}}}&Local Diffusion processes&DDPM, DDIM&OOD estimation with two modules: a ``branching'' module that predicts inside and outside OOD regions, and a ``fusion'' module that combines them into a unified output\\
\cline{2-5}
&\multirow{1}*{\shortstack{\cite{betti2024optimizing}}}&DHEaD (Hallucination Early Detection)&SD2& Combine cross-attention maps with a new metric, the predicted final image, to anticipate the final result using information from the early phases of generation\\
\cline{2-5}
&\multirow{1}*{\shortstack{\cite{Aithal2024Understanding}}}&Mode Interpolation&DDPM& Characterize the variance in the sample's trajectory during the final few backward sampling steps\\
\hline
\end{tabular}
\label{tab:hallucination_defense}
\end{table*}

\begin{figure}[tbhp]
\centering
\includegraphics[width=1.0\linewidth]{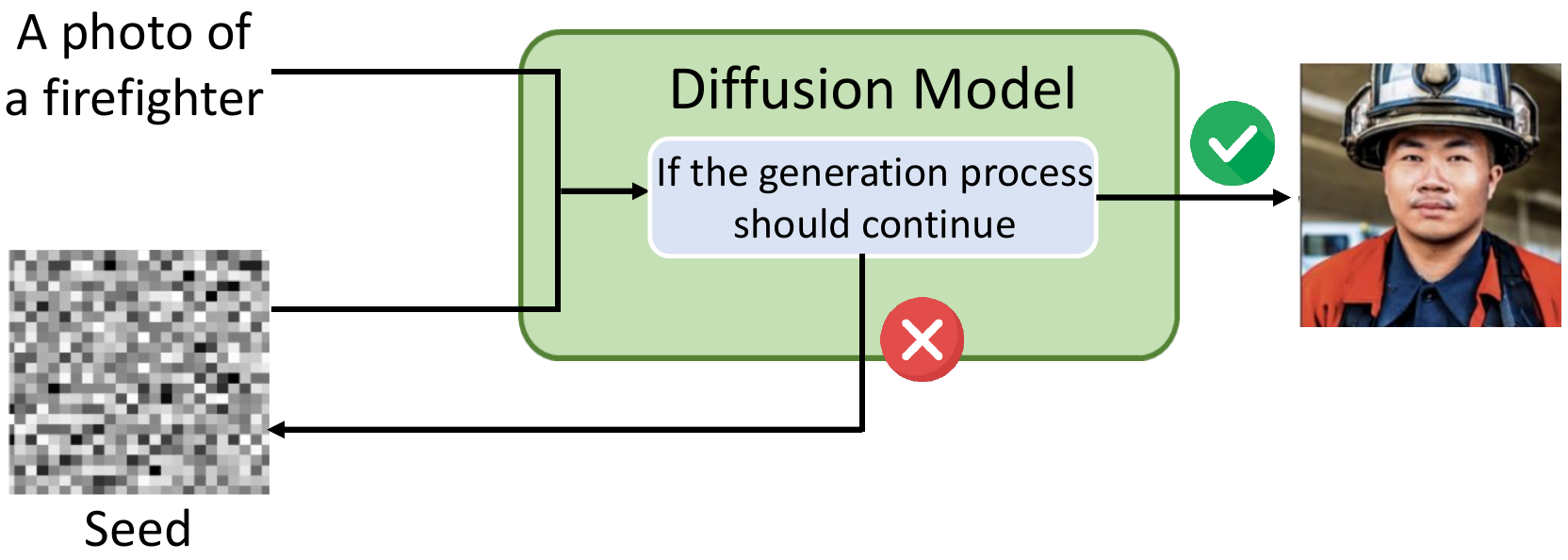}
\caption{Hallucination represents that DMs generate artifacts beyond real data characteristics.}
\label{fig:Hallucination}
\end{figure}
\subsection{Truthfulness}\label{subsec:truthfulness}
Truthfulness of DMs is inherently tied to the quality and composition of their training data. 
Biases, errors, or skewed representations in the data inevitably lead to false outputs, namely hallucinations. We summarize the hallucination issues and countermeasures in Tabs. \ref{tab:hallucination_issues} and \ref{tab:hallucination_defense}, and introduce them as follows.
\subsubsection{Hallucination}
\noindent\textbf{Risks:} DMs are known to generate hallucinatory samples outside real data characteristics \cite{Aithal2024Understanding, qin2024evaluating, betti2024optimizing, Kim2025Tackling}, as shown in Fig. \ref{fig:Hallucination}.
The work in \cite{qin2024evaluating} first introduced an LLM-based framework for scene-graph-guided question answering to assess T2I DMs and a novel evaluation dataset containing human-annotated hallucination scores for generated images, allowing systematic analysis of spurious content in DM outputs.
Furthermore, Aithal \textit{et al.} \cite{Aithal2024Understanding} revealed that DMs perform smooth interpolation between adjacent modes in the training data distribution, yet frequently generate samples lying outside the original distribution's support. 
This out-of-distribution generation capability leads to hallucinatory artifacts and synthetic features absent from real data.

\noindent\textbf{Countermeasures:} By systematically studying the reasons for the manifestation of hallucination in DMs, Aithal \textit{et al.} \cite{Aithal2024Understanding} demonstrated that DMs inherently recognize their out-of-support generations, as evidenced by high-variance trajectories during the final backward sampling steps, quantitatively capturing the hallucination phenomenon.
Kim \textit{et al.} \cite{Kim2025Tackling} further proposed a training-free framework that mitigates hallucinations in DMs through parallel local diffusion processes, consisting of: (1) out-of-distribution (OOD) estimation, followed by (2) a branching module that generates samples both within and beyond OOD regions, and (3) a fusion module that optimally combines these predictions.
Betti \textit{et al.} \cite{betti2024optimizing} proposed Hallucination Early Detection (HEaD), a framework that predicts generation outcomes by combining cross-attention analysis with the proposed Predicted Final Image indicator, ensuring the approach enables early hallucination detection by leveraging intermediate generation features available during initial denoising steps.
\section{Applications}

DMs can produce high-quality results even from simple inputs, making them useful in healthcare, media, robotics, and security. However, using these technologies brings certain risks. This section introduces specific examples of how DMs are used, what dangers they might bring, and ways to protect against these dangers.


\subsection{Creative Media and Content Generation}

Beyond basic image generation, DMs are widely used in animation, illustration, and design. They enable rapid prototyping and support creative exploration. However, privacy issues arise when training data includes personal or proprietary artwork, as shown in Subsection \ref{subsec:privacy}. The safety risks stem from potential misuse of the outputs for offensive or misleading purposes, as shown in Subsection \ref{subsec:safety}. Fairness can be impacted if certain cultures or styles are underrepresented, leading to biased results as shown in Subsection \ref{subsec:fairness}. Copyright disputes may also emerge, especially regarding the originality and ownership of generated content, as shown in Subsection \ref{subsec:copyright}. Effective strategies include dataset curation, content moderation, and clear usage policies.

\subsection{Medical Image Generation}
In the medical field, DMs can be used to generate synthetic medical images for training and testing diagnostic models~\cite{Wu_Ji_Fu_Xu_Jin_Xu_2024}. These synthetic datasets protect real patient data but can still raise privacy concerns if the generated outputs reflect identifiable patterns~\cite{zhao2025frequency}, as shown in Subsection \ref{subsec:privacy}. Robustness is vital, as medical neural networks are susceptible to adversarial attack~\cite{jeong2024uncertainty,Medghalchi_2025_CVPR}, resulting in medical accidents, as shown in Subsection \ref{subsec:robustness}. Fairness is also a concern, particularly if the model underperforms on data from underrepresented groups, such as those with long-tailed age distributions or gender biases, as demonstrated in Subsection \ref{subsec:fairness}. DP and balanced sampling methods are key solutions for ensuring secure and fair medical usage.

\subsection{Robotics and Autonomous Systems}
DMs contribute to the training of autonomous systems by creating synthetic scenarios for navigation and decision-making~\cite{wang2024drivedreamer,xu2025diffscene}, reducing costs and broadening coverage. However, simulation errors can pose safety hazards if carried over to real-world applications.
For instance, robustness is challenged by adversarial examples that can mislead perception systems~\cite{298066,xie2024advdiffuser}. Fairness issues arise when models trained in limited environments induced by biased DMs, fail to adapt to diverse cases in the real-world deployment. 
Mitigation involves secure training, adversarial testing, and the incorporation of diverse scenarios into model development.


\subsection{Security and Surveillance}

In security contexts, DMs are used to produce synthetic data for facial recognition and behavior analysis~\cite{liu2024adv, 10.1145/3703626} in the surveillance system. These applications pose significant privacy concerns, especially when synthetic faces closely resemble real individuals, as shown in Subsection \ref{subsec:privacy}. Robustness risks are associated with the potential misuse of generated data for intrusive surveillance~\cite{liu2024adv,10531291}, where imperceptible and transferable noise generated by DMs misleads surveillance system, as shown in Subsection \ref{subsec:robustness}. Fairness is critical, as biased training data may result in unequal treatment of different demographic groups~\cite{perera2025unbiased}, as shown in Subsection \ref{subsec:fairness}. Copyright issues may occur when real-world data is used without proper consent~\cite{Zhu_2024_CVPR}, as shown in Subsection \ref{subsec:copyright}. Responsible use demands anonymization methods, regulatory compliance, and ethical oversight.

These examples highlight how the practical use of DMs is closely tied to critical concerns in security and ethics. Understanding these risks in real applications helps guide the development of safer and fairer generative AI systems.


\section{Lessons Learned}
Through our study, we observed that many studies have explored the vulnerabilities of DMs. However, there are still many unexplored risks. In addition, defensive methods against these threats are incomplete. In the following, we outline a few limitations and potential future directions.
\subsection{Limitations}
In this subsection, we discuss several aspects of the limitations as follows.
\begin{itemize}[leftmargin=1em]
    \item \textbf{Strong assumptions:} A part of attacks consider the white-box scenario, such as jailbreak and adversarial attacks. However, these methods usually have poor performance when applied to commercial DMs that are inaccessible, such as data extractions \cite{Nicolas2023Extracting,wen2024detecting,chen2024extracting}. 
    Thus, it is necessary to explore the deep vulnerabilities of DMs. 
    \item \textbf{Theoretical understanding:} As introduced in Section \ref{sec:DMs}, the design of DMs relies on theoretical guidance, but most attack or defensive methods lack theoretical analysis or guarantee.
    The backdoor attacks need to modify the training process, i.e., inject transitions to diffuse adversarial targets into a biased Gaussian distribution, but they relied on the existing diffusion process without specific designs.
    For other attack and defensive methods, they are usually motivated by observations from experiments \cite{wen2024detecting}, which lack interpretability.
    \item \textbf{Threats are growing quickly:} 
    Attack strategies targeting DMs, such as prompt injection, data leakage, and adversarial examples, are developing rapidly. In contrast, defenses often emerge reactively. This highlights the urgent need for proactive security design, anticipating vulnerabilities before models are widely deployed.
    Another good alternative is to develop advanced training strategies that are capable of defending against evolving attack methods.
    \item \textbf{Unique vulnerabilities of DMs:} 
    DMs differ significantly from classification models or GANs due to their iterative denoising process, stochastic sampling, and high-dimensional latent spaces. As a result, generic threat models and defenses are often insufficient. New attack and defense frameworks are essential to address specific vulnerabilities of DMs.
    \item \textbf{Trade-offs between security and utility:} 
    Many defense mechanisms, such as DP, watermarking, and content filtering, introduce trade-offs in terms of generation quality, computational efficiency, or user experience. Future research should focus on quantifying and optimizing these trade-offs to balance robustness and usability in real-world applications.
    \item \textbf{Lack of standardized evaluation benchmarks:} 
    Current research lacks consistent benchmarks for evaluating security and privacy in DMs, particularly for qualitative concerns like fairness and copyright. This makes it difficult to compare different methods. 
\end{itemize}

\subsection{Future Directions}

To address the aforementioned limitations, we introduce the following directions.
\begin{itemize}[leftmargin=1em]
    \item \textbf{Benchmarks and evaluation:} Benchmarking involves defining standardized tasks, datasets, metrics, and attack methodologies to objectively compare the security and privacy performance against defense mechanisms in different DMs. Specifically, this work comprises the following parts: 1) Standardized datasets for attacks. Create publicly available datasets designed for specific attack types. 
    2) Attack benchmarks. Develop and maintain a suite of diverse, SOTA attack methods to rigorously test the robustness of DMs. 3) Defense evaluation. Evaluate the effectiveness of various defense mechanisms against these attacks, measuring their impact on both security and model utility. 
    4) Red teaming. Employ dedicated teams to actively probe and find vulnerabilities in DMs, simulating real-world adversaries. Agent-as-a-Judge \cite{zhuge2025agentasajudge} is a powerful and scalable approach to automate the ``red-teaming'' process, where you try to break a model's safety features to see how robust they are. 
    \item \textbf{Theoretical analysis on attacks and defenses:} 1) Information-theoretic lower bounds on privacy leakage. Can we establish theoretical limits on how much private information a DM inevitably reveals, regardless of the defense? This may involve analyzing mutual information between training data and model parameters/outputs. 2) Robustness to perturbations. How do small perturbations in the input or latent space propagate through the diffusion process? Can we analytically derive the ``robustness radius'' for DMs under specific noise schedules or model architectures? 3) Score function sensitivity. Analyze the sensitivity of the learned score function to input perturbations. A highly sensitive score function could indicate vulnerability to adversarial attacks. 4) Memorization analysis. Theoretically quantify the extent to which DMs ``memorize'' training data. This often involves concepts from statistical learning theory, such as generalization bounds and capacity measures~\cite{Li2023On}. 
    \item \textbf{Invisibility of attacks:} Many attacks aim to evade both human detection and defense mechanisms. 
    After all, an attack that is easily spotted is of little practical use. In DMs, researchers have developed sophisticated techniques to ensure the invisibility of their attacks. 
    One of the most effective methods for achieving invisibility is to operate in the latent space, rather than directly manipulating the pixels of an image or prompt. 
    Another technique that contributes to invisibility is the use of attention maps \cite{Wang2024Compositional}. Attention mechanism, a key component of the DMs, can be used to guide the attack to specific regions of an image, leaving the overall structure intact. This allows for more subtle and targeted manipulations. 
    \item \textbf{Comprehensive defenses against privacy risks:} Comprehensive defenses against privacy risks in DMs require a multi-faceted approach, addressing vulnerabilities at various stages of the model's life cycle, from data collection and training to deployment and inference, including: 1) Data-Centric Defenses (Protecting the Source); 2) Model-Training Defenses (Building a Secure Model); 3) Inference-Time and Post-Processing Defenses (Securing the Use). By combining these data, training, and inference-time strategies, developers can build a robust defense system that significantly mitigates the privacy risks associated with powerful DMs.
    \item \textbf{Safety alignment for DMs:} This involves designing mechanisms that prevent harmful, biased, or undesirable outputs while promoting a safe and ethical user experience. Key suggestions include: 1) Define Clear Safety Objectives. Before development, explicitly define what constitutes ``safe'' behavior. This includes identifying potential harms (e.g., generating hate speech, promoting self-harm, privacy violations, perpetuating stereotypes, creating disturbing content) and establishing guardrails against them. 2) Human-in-the-Loop (HITL) for Oversight. Implement a robust system for human review and intervention, especially during training and early deployment. This allows for real-time identification and correction of problematic outputs. 3) Transparency and Explainability. Design the DM to be as transparent as possible about its AI nature and limitations. If a DM is unable to fulfill a request due to safety concerns, it should ideally explain why. 
    4) User Consent and Control. Users should have clear control over their interactions and the ability to report or flag problematic content. Provide options to opt out of specific content or interactions. 
\end{itemize}

\section{Conclusions}
This paper presents a comprehensive survey on DMs from the safety, ethics, and trust perspectives.
We begin with a self-contained introduction to four fundamental frameworks of DMs: DDPMs, DDIMs, SGMs, and Score SDEs. 
Next, we discuss recent efforts on attack and defense methods in DMs across six primary aspects, i.e., privacy, robustness, safety, fairness, copyright, and truthfulness. 
A survey across four application domains, as well as an examination of risks, illustrates the necessity of exploring countermeasures in DMs. 
Finally, we conclude this paper with discussions on challenging issues for future research in this field.


%





\ifCLASSOPTIONcaptionsoff
  \newpage
\fi

\bibliographystyle{IEEEtran}
\bibliography{ref}

\end{document}